\documentclass[twocolumn]{aastex63}

\NewPageAfterKeywords

\graphicspath{{./}{figures/}}

\received{September 13, 2019}
\revised{October 11, 2019}
\accepted{October 14, 2019}
\submitjournal{ApJS}

\shorttitle{PFSS Modelling of PSP Encounter 1}
\shortauthors{Badman et al.}

\begin{document}
	
	\title{Magnetic connectivity of the ecliptic plane within 0.5 AU : PFSS modeling of the first PSP encounter }

	\correspondingauthor{Samuel T. Badman}
	\email{samuel\_badman@berkeley.edu}
	
	\author[0000-0002-6145-436X]{Samuel T. Badman}
	\affil{Physics Department, University of California, Berkeley, CA 94720-7300, USA}
	\affil{Space Sciences Laboratory, University of California, Berkeley, CA 94720-7450, USA}
	
	\author[0000-0002-1989-3596]{Stuart D. Bale}
	\affil{Physics Department, University of California, Berkeley, CA 94720-7300, USA}
	\affil{Space Sciences Laboratory, University of California, Berkeley, CA 94720-7450, USA}
	\affil{The Blackett Laboratory, Imperial College London, London, SW7 2AZ, UK}
	
	\author[0000-0002-2587-1342]{Juan C. Mart\'inez Oliveros}
	\affil{Space Sciences Laboratory, University of California, Berkeley, CA 94720-7450, USA}
	
	\author[0000-0002-4440-7166]{Olga Panasenco}
	\affil{Advanced Heliophysics, Pasadena, CA 91106, USA}

	\author[0000-0002-2381-3106]{Marco Velli}
	\affil{EPSS, UCLA, Los Angeles, CA 90095, USA}
	
	\author[0000-0002-1365-1908]{David Stansby}
	\affil{Mullard Space Science Laboratory, University College London, Holmbury St. Mary, Surrey RH5 6NT, UK}
	\affil{Department of Physics, Imperial College London, London SW7 2AZ, UK}
	
	\author[0000-0002-8203-4794]{Juan C. Buitrago-Casas} 
	\affil{Physics Department, University of California, Berkeley, CA 94720-7300, USA} 
	\affil{Space Sciences Laboratory, University of California, Berkeley, CA 94720-7450, USA} 
	
	\author[0000-0002-2916-3837]{Victor R\'eville}
	\affil{EPSS, UCLA, Los Angeles, CA 90095, USA}

	\author[0000-0002-0675-7907]{J. W. Bonnell}
	\affil{Space Sciences Laboratory, University of California, Berkeley, CA 94720-7450, USA}

	\author[0000-0002-3520-4041]{Anthony W. Case}
	\affil{Smithsonian Astrophysical Observatory, Cambridge, MA 02138, USA}

	\author[0000-0002-4401-0943]{Thierry {Dudok de Wit}}
	\affil{LPC2E, CNRS and University of Orl\'eans, Orl\'eans, France}

	\author[0000-0003-0420-3633]{Keith Goetz}
	\affiliation{School of Physics and Astronomy, University of Minnesota,
		Minneapolis, MN 55455, USA}

	\author[0000-0002-6938-0166]{Peter R. Harvey}
	\affil{Space Sciences Laboratory, University of California, Berkeley, CA
		94720-7450, USA}

	\author[0000-0002-7077-930X]{J. C. Kasper}
	\affiliation{Climate and Space Sciences and Engineering, University of Michigan, Ann Arbor, MI 48109, USA}
	\affiliation{Smithsonian Astrophysical Observatory, Cambridge, MA 02138, USA}

	\author[0000-0001-6095-2490]{Kelly E. Korreck}
	\affil{Smithsonian Astrophysical Observatory, Cambridge, MA 02138, USA}

	\author[0000-0001-5030-6030]{Davin E. Larson}
	\affil{Space Sciences Laboratory, University of California, Berkeley, CA 94720-7450, USA}

	\author[0000-0002-0396-0547]{Roberto Livi}
	\affil{Space Sciences Laboratory, University of California, Berkeley, CA 94720-7450, USA}

	\author[0000-0003-3112-4201]{Robert J. MacDowall}
	\affil{Solar System Exploration Division, NASA/Goddard Space Flight Center, Greenbelt, MD, 20771}

	\author[0000-0003-1191-1558]{David M. Malaspina}
	\affil{Laboratory for Atmospheric and Space Physics, University of Colorado, Boulder, CO 80303, USA}
	
	\author[0000-0002-1573-7457]{Marc Pulupa}
	\affil{Space Sciences Laboratory, University of California, Berkeley, CA 94720-7450, USA}
			
	\author[0000-0002-7728-0085]{Michael L. Stevens}
	\affil{Smithsonian Astrophysical Observatory, Cambridge, MA 02138, USA}
		
	\author[0000-0002-7287-5098]{Phyllis L. Whittlesey}
	\affil{Space Sciences Laboratory, University of California, Berkeley, CA 94720-7450, USA}

	
	
	\begin{abstract}
	We compare magnetic field measurements taken by the FIELDS instrument on Parker Solar Probe (PSP) during it's first solar encounter to predictions obtained by Potential Field Source Surface (PFSS) modeling. Ballistic propagation is used to connect the spacecraft to the source surface. Despite the simplicity of the model, our results show striking agreement with PSP's first observations of the heliospheric magnetic field from $\sim$0.5 AU (107.5 $R_\odot$) down to 0.16 AU (35.7 $R_\odot$). Further, we show the robustness of the agreement is improved both by allowing the photospheric input to the model to vary in time, and by advecting the field from PSP down to the PFSS model domain using in situ PSP/SWEAP measurements of the solar wind speed instead of assuming it to be constant with longitude and latitude. We also explore the source surface height parameter ($R_{SS}$) to the PFSS model finding that an extraordinarily low source surface height ($1.3-1.5R_\odot$) predicts observed small scale polarity inversions which are otherwise washed out with regular modeling parameters. Finally, we extract field line traces from these models. By overlaying these on EUV images we observe magnetic connectivity to various equatorial and mid-latitude coronal holes indicating plausible magnetic footpoints and offering context for future discussions of sources of the solar wind measured by PSP. 	
	\end{abstract}
	
	

	\section{Introduction}
	
	Parker Solar Probe \citep[PSP;][]{Fox2016} is a NASA mission intended to revolutionize our understanding of the solar corona by becoming the first spacecraft to measure it's outer layers in situ. The fundamental science objectives are to (1) Trace the flow of energy that heats and accelerates the solar corona and solar wind; (2) Determine the structure and dynamics of the plasma and magnetic fields at the sources of the solar wind; and (3) Explore mechanisms that accelerate and transport energetic particles \citep{Fox2016}. 

	Central to it's science goals is PSP's record-breaking orbit. PSP launched on August 12 2018 and, after it's first Venus gravity assist, entered into the closest-grazing heliocentric orbit ever reached by an artificial satellite. On November 6 2018 PSP completed it's first perihelion pass at 35.7$R_\odot$ from the Sun traveling at almost 100 km s$^{-1}$. Future Venus gravity assists will eventually asymptote these numbers to 9.86 $R_\odot$ closest approach at over 200 km s$^{-1}$ in December 2024. A unique outcome of achieving this rapid orbital velocity is that PSP briefly reached a greater angular velocity than the Sun's equator.  This means it moves very slowly relative to the local corotating magnetic structure and samples the same  solar meridians multiple times in the same orbit (Figure \ref{fig1}C).
	
	PSP carries a suite of four scientific instruments. The electromagnetic fields investigation \citep[FIELDS;][]{Bale2016} probes in situ electric and magnetic fields and plasma waves, the spacecraft potential, quasithermal noise and low frequency radio waves. The Solar Wind Electrons Alphas and Protons investigation \citep[SWEAP;][]{Kasper2016} provides distribution functions and density, velocity and temperature moments of the most abundant species in the solar wind. The Integrated Science Investigation of the Sun \citep[IS$\odot$IS;][]{McComas2016} observes energetic electrons, protons and heavy ions from 10s of keV to 100 MeV.The Wide-Field Imager for Solar Probe Plus \citep[WISPR;][]{Vourlidas2016} is a white light imager which observes structures in the solar wind, such as shocks, approaching and passing the spacecraft. In this work we utilize FIELDS DC magnetic field data and SWEAP proton velocity moments.
	
	A major source of contextual information for spacecraft in situ measurements of the solar wind comes from global coronal and heliospheric modeling. Modeling techniques of varying complexity \citep[see e.g. review by][]{Wiegelmann2017} have been developed using historical measurements (both remote and in situ) as boundary conditions. PSP provides unique constraints on such models given it is sampling entirely new regions of the heliosphere. It is therefore of great interest to compare PSP observations to these models both to contextualize the measurements and to improve the models themselves.
	
	In this work we compare PSP magnetic field observations with predictions made using the widely used Potential Field Source Surface (PFSS) model \citep{Altschuler1969, Schatten1969, Hoeksema1984, Wang1992}. PFSS employs two key assumptions: (1) A current-free corona, which is a special case of force free models which require an assumption of very low plasma beta (meaning magnetic pressure dominates over thermal pressure). (2) A spherical source surface of heliocentric radius $R_{SS}$ at which field lines are enforced to be radial, simulating the role of the solar wind in opening these field lines out to interplanetary space. Despite these assumptions PFSS compares well to more modern magnetohydrodynamic (MHD) models \citep[e.g.][]{Riley2006} and is widely used due to it's computational simplicity and associated high resolution. 
	
    With zero currents the magnetic field, \textbf{B},  in PFSS may be expressed as a scalar potential, $\Phi_B$, such that $\textbf{B} = -\nabla \Phi_B$. By the ``no-monopole'' Maxwell equation, this potential must obey the Laplace equation, $\nabla^2 \Phi_B = 0$, for which solutions are very well understood. PFSS solves for the field in an annular volume of radial extent $1 R_\odot \leq r \leq R_{SS}$. Boundary conditions are the measured radial magnetic field at the photosphere ($1 R_\odot$) and the requirement that the tangential components of \textbf{B} vanish at $R_{SS}$. This allows the problem to be uniquely solved as a spherical harmonic decomposition, resulting in a full 3D magnetic field model between the photosphere and source surface. The solution is steady state and represents a low energy bound on more general force free models, \citep{Regnier2013}.
		
	PFSS models have historically been used to predict magnetic polarity at 1AU \citep[e.g][]{Hoeksema1984}, coronal structure observed during solar eclipses \citep[e.g.][]{Altschuler1969}, and identifying open field line regions associated with coronal holes \citep[e.g.][]{Wang2019}. In addition, even though PFSS only models the magnetic field directly, \citet{Wang1990} showed an inverse correlation between the divergence rate of PFSS field lines with observed solar wind speed at 1AU, indicating that coronal magnetic field topology plays an important role in the acceleration of the solar wind. This observation has since been refined by \citet{Arge2000,Arge2003, Arge2004} into the modern WSA model which assimilates PFSS and a Schatten current sheet model \citep{Schatten1972}, and is currently used operationally in space weather predictions and hosted by NASA's Coordinated Community Modeling Center (CCMC, \url{http://ccmc.gsfc.nasa.gov}).

	Here, we report our results obtained from the use of a simple PFSS model and a ballistic propagation model \citep{Nolte1973} to connect the spacecraft to the PFSS model domain, and use this to explain features of large scale magnetic structure observed in PSP's first solar encounter. In Section \ref{PSP_DATA} we introduce the data taken by PSP used in this work as well as using it to extrapolate magnetic polarity structure out to 1AU and compare to measurements by the Magnetic Field Investigation \citep[MFI;][]{Lepping1995} on board the Wind spacecraft \citep{Harten1995}. In section \ref{MODELING_METHOD} the implementation of PFSS modeling and the procedure to connect those results to the in situ measurements of PSP are described. Section \ref{RESULTS} lists the major results from this work: 1) General successful prediction of in situ timeseries measurements. 2) Improvements to modeling through time evolving magnetospheric inputs and use of PSP/SWEAP solar wind velocity measurements. 3) Recovery of smaller scale structure consistent with measurements when the source surface height parameter ($R_{SS}$) of the PFSS model is dramatically lowered. 4) Identification of mid-latitude and equatorial coronal holes as potential sources of the solar wind PSP measured in it's first encounter. We conclude in section \ref{DISCUSSION} with a discussion of the results and interpretation with particular attention to address the limitations of this modeling method in light of it's simplifications, and reference concurrent and future modeling work.
	
	\section{PSP Data}
	\label{PSP_DATA}
	
	\subsection{Timeseries and spatial distribution.}
	
	We begin by presenting the data used in this work from the FIELDS and SWEAP instruments on board PSP measuring during the first solar encounter (E1) from 2018-10-15 to 2018-11-30. From FIELDS we use measurements of the radial component of the magnetic field ($B_r$) and from SWEAP we use the radial component of the proton velocity moment ($V_{SW}$). Since the focus of this work is the large scale solar wind structure, we first pre-process this data to remove transients and rapid fluctuations such as the newly observed $\delta B/B\sim 1$ magnetic  ``switchbacks'' and velocity spikes (see e.g. \citet{Bale19,Kasper19,DudokdeWit2020,Horbury2020}. To do this we bin the full cadence data into hourly segments, generate a histogram of the data in each bin and take the modal value (the value corresponding to the peak of the histogram). 

	\begin{figure}	
		\centering
		\plotone{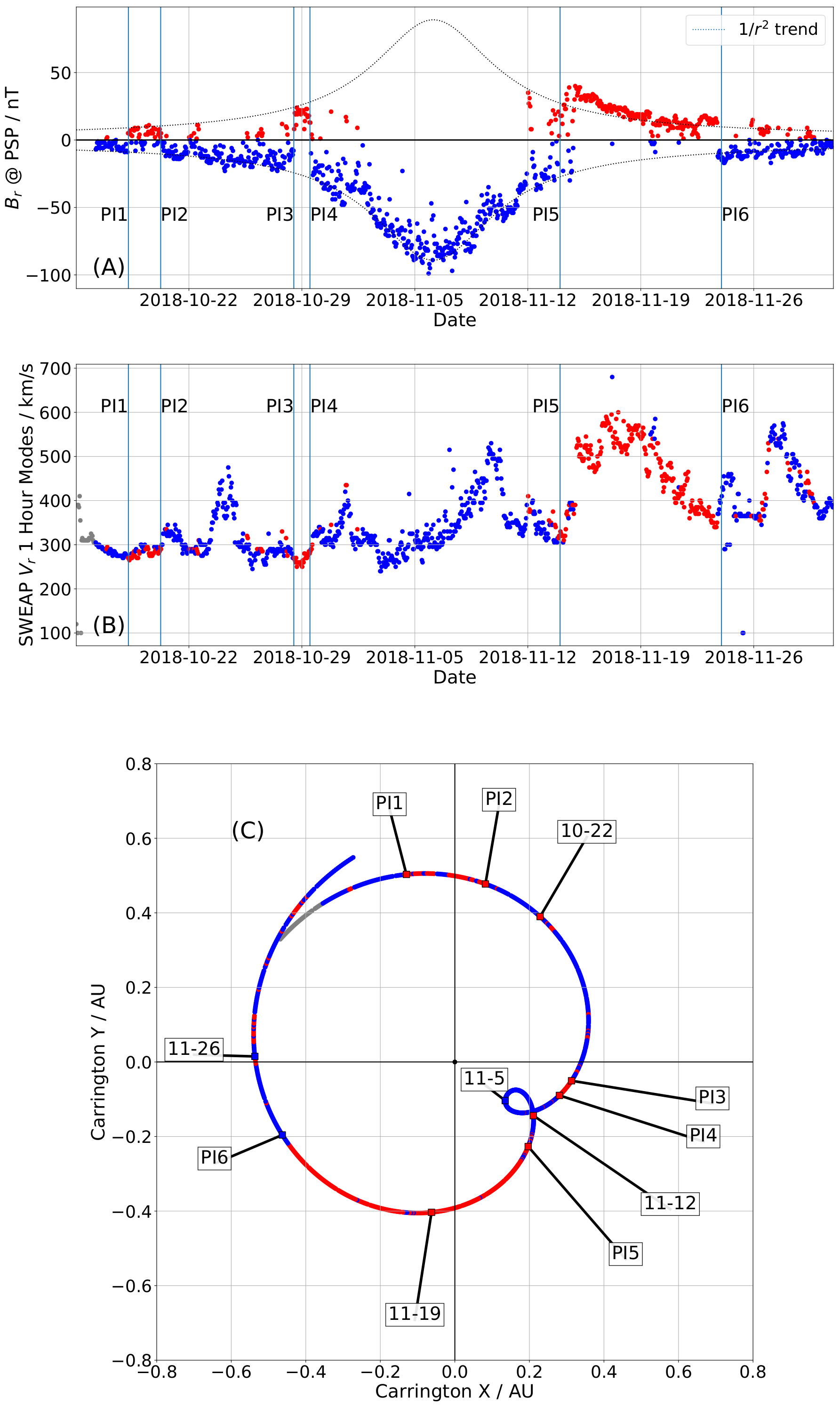}
		\caption{\textbf{Exposition of 1 hour modes (see main text) of PSP data taken during Encounter 1 (E1) from 2018-10-15 to 2018-11-30.} Perihelion occurred on 2018-11-6. Panel (A): PSP/FIELDS radial magnetic field ($B_r$) vs. time colored by magnetic polarity: Positive (red) means radially outwards, while negative (blue) means radially inwards. A $1/r^2$ dotted trend line shows the zeroth order behavior. Panel (B): PSP/SWEAP radial proton bulk velocity ($V_{SW}$) vs. time colored by measured magnetic polarity.  Panel (C): The E1 orbital trajectory of PSP plotted in Carrington (solar-corotating) coordinates. The trajectory is colored by the measured magnetic polarity, demonstrating the apparent spatial structure of the magnetic sectors sampled.}
		\label{fig1}
	\end{figure}
	
	The resulting time series are shown in figure \ref{fig1}. $B_r$ is shown in panel (A), in which we identify a $1/r^2$ overall trend (dotted line) and colorize by polarity with the convention of red for radially outwards ($B_r > 0$) and blue for radially inwards ($B_r < 0$). This convention will be followed in all plots in this paper. The time of perihelion at $35.7 R_\odot$ is easily identified by the occurence of maximum field strength on 2018-11-6. The time series shows generally negative polarity indicating the orbital trajectory was mostly on the southward side of the heliospheric current sheet (HCS), and a number of clear multi-hour excursions into positive polarity on 2018-10-20 (bounded by PI1 and PI2), 2018-10-29 (bounded by PI3 and PI4) and from 2018-11-14 (PI5) through to 2018-11-23 (PI6). A detailed look at the nature of these inversions as PSP crossed the HCS is given in \citet{Szabo2020}. Positive spikes on 2018-10-23 and 2018-11-13 are attributed to small coronal mass ejections (CMEs) as discussed in \citet{McComas2019, Korreck2020, Nieves-Chinchilla2020, Giacalone2020, Mitchell2020}. 
	
	$V_{SW}$ is shown in panel (B) and colorized by the magnetic polarity for ease of comparison with the $B_r$ time series. We observe a mostly slow wind ($<$500 km s$^{-1}$) which is generally uncorrelated with the magnetic polarity inversions, except for PI5 which is coincident with a sudden transition to a moderately fast wind stream peaking at about 600km s$^{-1}$.
	
	Panel (C) of figure \ref{fig1} demonstrates the spatial distribution of magnetic polarities observed during PSP's Encounter 1. We plot the spacecraft trajectory (projected on to the solar equatorial plane) in Carrington coordinates (i.e. corotating with the Sun's equator) and color it according the polarity of $B_r$. In this co-rotating frame, PSP starts in the upper left quadrant and tracks clockwise (i.e. retrograde). As it's radial distance from the sun decreases, it's angular velocity catches up to that of the Sun, eventually reaching co-rotation prior to perihelion, it briefly rotates faster than the Sun before transitioning back to sub-corotational speeds as it climbs to higher altitudes. This results in the small loop centered near the 2018-11-5 date label. It should also be noted that this changing angular speed means the rate of change of longitude of PSP varied dramatically; two out of the six week interval shown were spent between the longitudes of the co-rotation loop. The extent and location of the positive polarity inversions in the time series are identified as the red regions of the trajectory. Clearly the PI1-PI2 and PI3-PI4 intervals occurred over a small range of solar longitudes ($< 10$ degrees), while the interval between PI5 and PI6 was protracted over a much larger region, spanning about 90 degrees of solar longitude.

	\subsection{Parker spiral stream structure and comparison to 1AU Measurements}
	\label{pspiral_extrap}
	
	In the previous section, we have shown the in situ data of $B_r$ and $V_{SW}$ measurements and it's orbital context. The orbital context allows us to directly assign solar wind parameters along a narrow path through a complex 3D medium. In this section we seek to extrapolate these measurements to infer a solar wind stream structure to connect this data out to 1AU and compare it to Wind/MFI observations of the magnetic field polarity.

	\begin{figure*}
		\centering
		\plotone{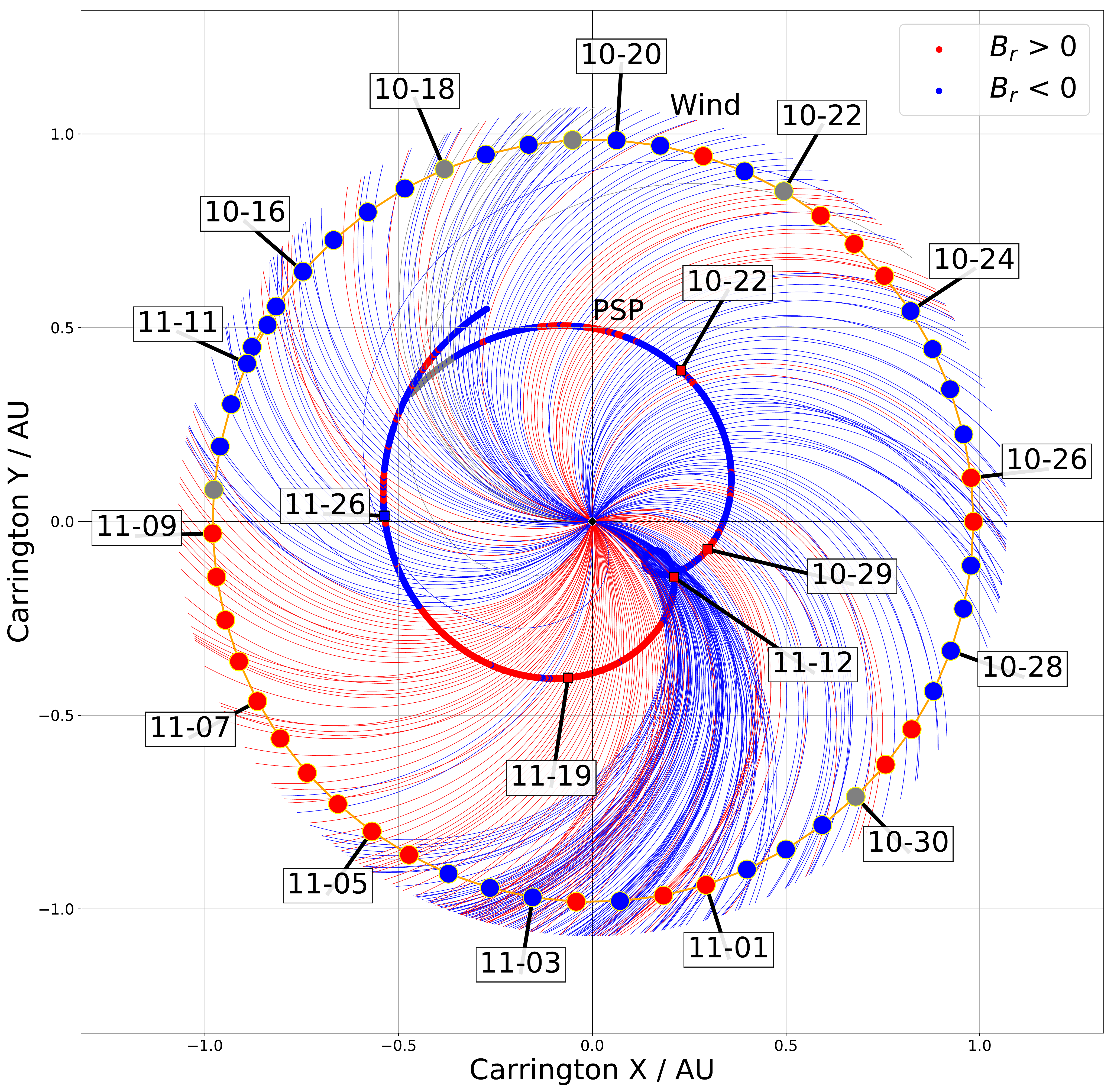}
		\caption{\textbf{Magnetic polarity sector structure implied by PSP extrapolated to 1AU.} Parker spiral field lines are initialized by the SWEAP $V_{SW}$ measurements and colored by the FIELDS $B_r$ polarity. These are propagated out to 1 AU. Measurements of the magnetic polarity by Wind/MFI are shown at a 12 hourly cadence. For each time, positive polarity is designated if the field is $>7/12$ positive for $\pm6$ hours of this time and similar for negative, otherwise the polarity is designated mixed and colored grey. As in figure \ref{fig1} (C), the coordinates corotate with the Sun.}
		\label{fig2}
	\end{figure*}
	
	To accomplish this, we use the Parker spiral \citep{Parker1958} approximation in which we assume each point along the PSP trajectory is threaded by a Parker spiral field line with a curvature determined by the in situ solar wind speed, giving the spiral the following 2D parameterization of longitude and radius ($\phi,r$): 
	
	\begin{equation}
		\label{psiral_eqn}
		\phi(r) = \phi_0 - \frac{\Omega_\odot}{V_{SW}}(r-r_0)
	\end{equation}
	
	where $\phi_0,r_0$ are the longitude and radial distance of PSP at the time of the $V_{SW}$ measurement. $\Omega_\odot$ is the solar rotation rate which we calculate from the equatorial rotation period of 24.47 days, assuming the latitudinal offset of PSP ($< 5$ degrees) is small enough to not consider differential rotation. This equation shows the qualitative dependence of the spiral field lines of solar wind speed: faster $V_{SW}$ gives straighter field lines (smaller $d\phi/dr$), slower wind speed produces more curved field lines (larger $d\phi/dr$).

	Each spiral field line is colored by the PSP measured polarity. The result is plotted in figure \ref{fig2}, again in Carrington coordinates. The date-labeled PSP trajectory is shown in the context of the stream structure out to 1AU where we plot the trajectory of the Wind spacecraft, located at the Earth L1 point, for a similar date range. At a 12 hour cadence, we plot the polarity of the radial magnetic field measured by the Wind/MFI instrument. Guided by a similar convention from \citet{Hoeksema1983}, to assign this 12 hour interval a single polarity, we take all measurements at a minute cadence from $\pm$ 6 hours of the measurement time and assign a positive polarity (red) if more than 7/12 of all data values are positive, and negative polarity (blue) if more than 7/12 of all data values are negative. If neither of these criteria are satisfied, the field is designated ``mixed'' for that interval and colored grey. 
	
	The choice of generating field lines at a constant time interval means that as the relative angular velocity of the spacecraft gets smaller, the field lines appear to bunch together more. We emphasize this is a pure sampling effect and does not indicate anything physical about the field. It is further compounded by the corotation loop which means several field lines are generated at the same Carrington longitude.
	
	It is important to note that PSP and Wind sampled the same solar longitudes at different times due to the differing orbits. In displaying the picture in figure 2, we are assuming the structure shown is essentially fixed in time as the spacecraft travel through it making measurements. Even so we see the dominant features in the PSP data bear out at 1AU using this simple picture. The PI1-PI2 interval connects via the spiral to positive polarity at 1AU. The PI3-PI4 interval merges into slowing wind which also mixes the negative polarity at perihelion with the onset of the positive polarity and fast wind during PI5.Although, as noted above, this region of the plot is made more complicated by the perihelion loop, the boundaries where field lines overlap further out than PSP represent likely locations where the Parker Spiral assumption breaks down and field lines would bend into compressions or rarefactions. The more complex picture of  interacting streams corresponds to a greater mix of positive and negative polarity measured as Wind sampled this region. The continuous period of positive polarity between PI5 and PI6 at PSP is also seen by Wind, with the Parker spiral very accurately predicting the transition from positive back to negative (PI6) by Wind on 2018-11-10, well in advance of PSP traversing the same structure on 2018-11-23.
	
	This good correspondence between PSP and 1AU measurements suggests the polarity inversions observed at PSP are relatively stable large scale features which must have an origin in open flux emerging from the solar corona. In addition, the implementation of the Parker spiral model lays the groundwork for the dicussion in the next section on connecting PSP measurements inwards to the corona where they can be compared to modeling results. As mentioned above, we also note that the stream structure implies magnetic pile-ups (and therefore divergence from the Parker spiral picture) further out than PSP, suggesting the locations of stream interaction regions (SIRs) which are studied in \citet{Allen2020} and \citet{Cohen2020}. However, tracing the field lines inwards, the streams should interact and distort each other much less.
	
	\section{Modeling Method}
	\label{MODELING_METHOD}
	
\begin{figure*}
		\centering
		\plotone{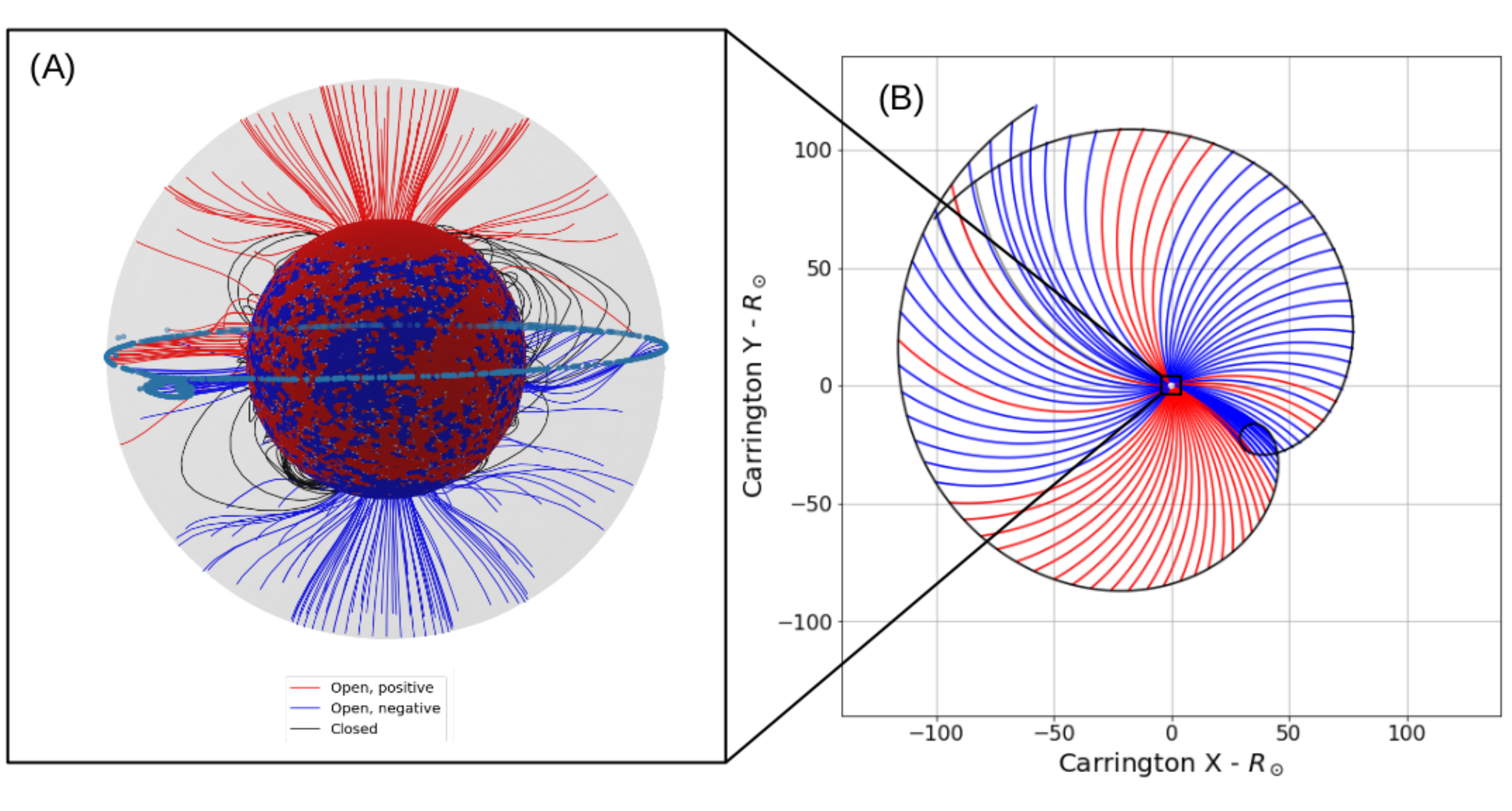}
		\caption{\textbf{Schematic of process to connect PSP Measurements to PFSS Modelling.} Panel (A) : PFSS model output from \textit{pfsspy}. The synoptic magnetogram input is shown as the photospheric (inner) boundary.  The model domain is bounded at at the exterior by the source surface (grey surface). Field lines initialized by a uniform grid at the photosphere are shown. Panel (B) : The outer boundary of the model is connected to the orbital position of PSP via an ideal Parker spiral magnetic field line. With some choice of solar wind speed, this maps the PSP trajectory to a locus of latitudes and longitudes at the source surface. This is illustrated by the near equatorial blue data points on the source surface in Panel (A).  }
		\label{fig3}
	\end{figure*}

	Having introduced the PSP data from E1, it's spatial context and salient features, we now introduce our PFSS modeling procedure and method of producing resulting time series predictions at PSP. This procedure is schematically illustrated in  Figure \ref{fig3}, and derives from \cite{Stansby2019b}.
	
	\subsection{PFSS Implementation} 
	
	The PFSS model is run in python using the open source \textit{pfsspy} package \citep{Yeates2018,Stansby2019a}. This code is available online and is very flexible, fast and easy to use. As input, it takes synoptic maps of the radial magnetic field at the photosphere and a value for the source surface height parameter $R_{SS}$. From these data it produces a full 3D magnetic field within the modeled volume, as well as a utility to trace individual magnetic field lines through the model solution. This input/output mapping is illustrated in Figure \ref{fig3} (A) which shows selected 3D field lines produced by \textit{pfsspy}, the extent of the model and the photospheric map which seeds the model. The grey surface illustrates the spherical source surface at which the field is constrained to be radial. As shown by the coloring of the field lines, they may either be open (red or blue depending on polarity) or closed (black). Open field lines are those which by definition intersect the source surface. Where they connect to the inner boundary of the model they indicate the probable locations of coronal holes (see Section \ref{OPEN_FIELDLINES}). During solar minimum, most open field lines emerge from large polar coronal holes. 
	 
	For the input magnetogram, there are a number of possible sources of data. In this work we initally considered the Global Oscillation Network Group \citep[GONG;][]{Harvey1996} zero corrected data product \citep{Clark2003}, and the Heliospheric Magnetic Imager \citep[HMI;][]{Scherrer2012} vector magnetogram data product \citep{Hoeksema2014}. GONG is measured from a network of ground based observatories and is operationally certified as input to a number of space weather prediction models. HMI, an instrument on board the Solar Dynamics Observatory \citep[SDO;][]{Pesnell2012}, is higher resolution and does not suffer atmospheric effects. Both of these have the limitation that they rely purely on observations and so cannot account for evolution on the far side of the Sun until that part of the Sun rotates into view. With this in mind, we have also used the Air Force Data Assimilative Photospheric Flux Transport (ADAPT) modeled magnetogram \citep{Arge2010} evaluated with GONG input, and the DeRosa/LMSAL modeled magnetogram \citep{Schrijver2003} (based on HMI) to compare results.  ADAPT and LMSAL make use of surface flux transport models into which new observations are assimilated. This procedure therefore models the far side evolution, implying a more accurate global picture of the photosphere. In practice, on discriminating between PFSS outputs from different magnetogram inputs using PSP data, we find little impact on our conclusions. We find GONG maps produce smooth predictions combining maps from one day to the next (section \ref{RESULTS1}), and require no pre-processing. While this smoothness may be a product of low resolution and atmospheric effects, it results in good clarity in displaying the features discussed in section \ref{RESULTS} without changing the conclusions. ADAPT maps resulted in very similar predictions but with some small fluctuation in the flux strength prediction from one day to the next which can be interpreted as model uncertainty. HMI included some missing days in the magnetogram record and does not include reconstruction of unobserved polar regions. The DeRosa/LMSAL model reconstructs the polar region and produces very similar predictions to the other models at lower source surface heights. For source surfaces much higher than $2.0R_\odot$ some deviation from the observations and other models takes place such as predicting constant positive field prior to October 29 (see Appendix A). This adds to the evidence we build in the results below that taking lower source surface heights in general is necessary for the best agreement between PFSS modeling and the observations.
	
	In addition, in certain parts of this work, we use a model from a single date to represent the entire encounter (figures \ref{fig6}C, \ref{fig7}, \ref{fig8}A). The extra model evolution of ADAPT or the DeRosa model actually makes this presentation difficult since times earlier than the model evaluation have changed significantly and no longer agree with what PSP measured at that time. For GONG, longitudes earlier in time are frozen after they go out of view and due to fortuitous orbital alignment the ``older'' parts of the model agree better with corresponding PSP measurements. For this reason, and in absence of any strong effects on conclusions, PFSS model results shown \textbf{in this paper} use GONG magnetograms unless otherwise stated. Further discussion of this choice, and comparisons of timeseries predictions using different magnetograms are included in Appendix \ref{AppA}.

	\subsection{Ballistic Propagation Model}
	
	PFSS only models the coronal magnetic field out to the source surface at a couple of solar radii. PSP on the other hand made in situ measurements at radii down to a minimum of 35.7$R_\odot$ during it's first encounter. To connect the model domain outwards to PSP's orbit we use ballistic propagation. Proposed by \cite{Nolte1973}, this technique assumes an arbitrary point in the heliosphere can be connected inwards to the corona by an ideal Parker spiral field line. The curvature of the spiral is driven by the solar wind speed measured out in the heliosphere. This model implicitly assumes that this measured wind speed (and therefore spiral curvature) remains constant all the way down into the corona. While this is not an accurate picture of the real solar wind, the dominant correction to this (acceleration of the solar wind) is counterbalanced by the effect of corotation meaning that the real footpoint of the field line is in fact close to where the ballistic propagation assumption puts it. \cite{Nolte1973} conclude the error in longitude of this method is within $10$ degrees. The implications of this error are discussed in section \ref{timeseries_rss}.
	
	Much like in section \ref{pspiral_extrap}, we assign a series of Parker spiral field lines to the orbit of PSP but this time propagate them \textit{inwards} to derive a longitude and latitude on the source surface to which each position along the PSP trajectory is connected to. The spiral field lines initialized from the PSP orbit are show in figure \ref{fig2}(B) and the resulting locus of coordinates on the source surface are indicated by the near equatorial blue scatter points in panel (A).  At each of these points on the source surface $B_r$ is obtained from the model and then multiplied by the value $r^2_{SS}/r^2_{PSP}$ at that time to project this model field out to PSP. Finally, since mean total unsigned flux between magnetograms generally do not agree with each other, rather than assume one particular magnetogram is correctly normalized, we scale our results by a constant factor so that their peak magnitude is equal to the peak magnitude measured by PSP at perihelion (figure \ref{fig1}A). As shown in figure \ref{fig6} and discussed in section \ref{DISCUSSION}, this factor is a function of source surface height and ranges from O(10) to order unity over the range of source surface heights considered in this work.

	\section{Results}
	\label{RESULTS}
	
	We now present the results of comparison of the above modeling to the PSP data.
	
	\subsection{Time series prediction and comparison}
	\label{RESULTS1}
		
\begin{figure*}
		\centering
		\plotone{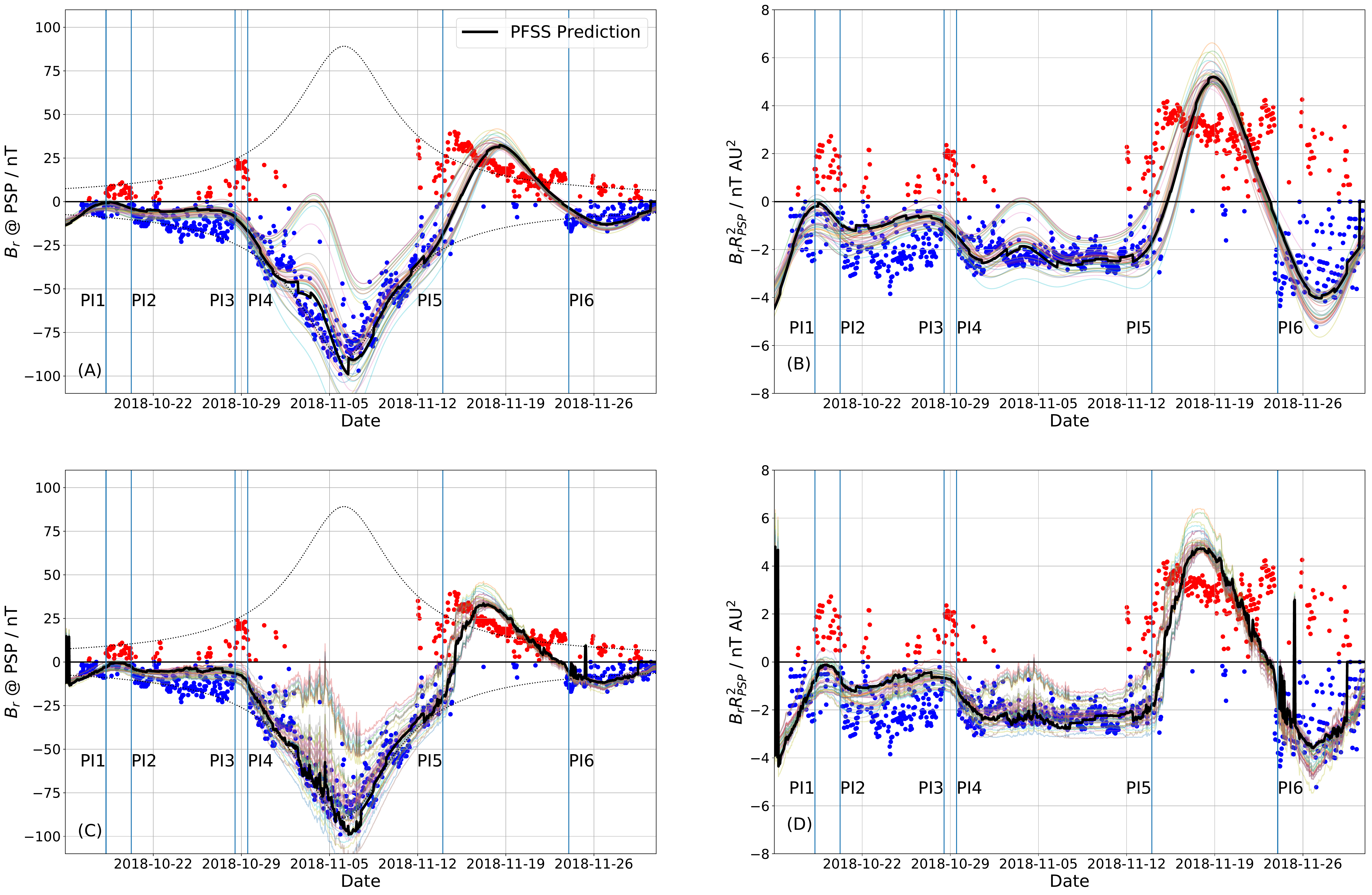}
		\caption{\textbf{Comparison of PFSS Predictions of $B_r$ with observations.}. The left hand column (A,C) shows predictions and measurements in situ at PSP, while the right hand column (B,D) scales out the $1/r^2$ dependence to compare predictions near the source surface. The top row (A,B) shows predictions made assuming a constant wind speed of $360 \text{km s}^{-1}$, while the bottom (C,D) row shows the results of using the SWEAP $V_{SW}$ measurements to connect the source surface to PSP. The faint colored lines in each plot indicate an ensemble of predictions made by using updating magnetograms at a cadence of 3 days. The solid black line indicates a synthesis of these predictions by combining segments from each prediction $\pm$1.5 days from the date of that photospheric map. Models all use GONG data and a source surface height of $R_{ss} = 2.0 R_\odot$}
		\label{fig4}
	\end{figure*}

	In figure \ref{fig4}, we overlay time series predictions using GONG magnetograms and a source surface height $R_{SS} = 2.0 R_\odot$ (the choice of this value is discussed in section \ref{SS}). The model results are plotted as lines while the $B_r$ data are plotted unchanged from figure \ref{fig1}(A) as a scatterplot colored by polarity.  We compare a number of modeling variations across the four panels:
	
	\begin{itemize}
	\item  In each panel a number of modeling realizations are plotted as faint colored lines. Each realization is generated using a GONG magnetogram from a different date spaced apart by 3 days over the encounter. We overplot a ``time-integrated'' model as a thick solid black line where these individual realizations are combined together: for each faint line only the data from $\pm$1.5 days from the date of the relevant magnetogram are used and stitched together sequentially. In terms of scaling, the models are stitched together first and then multiplied by a constant factor to match the peak measured $B_r$. For panels A and B this value is 6.90 while for panels C and D it is 6.73.
	
	\item From left to right (A,C vs B,D) we multiply the prediction and data by $r_{PSP}^2$ to compare the models and data without the $1/r^2$ scaling which dominates the overall shape of the time series in the raw data.
	
	\item We demonstrate the impact of using the measured solar wind velocity (bottom panels, C and D)  when generating Parker spirals to connect PSP to the source surface vs just assuming a constant slow solar wind speed of 360 km s$^{-1}$ (top panels, A and B).   
	\end{itemize}

	We observe an overall good agreement with a generally negative polarity field, dominantly varying with $1/r^2$ as predicted by the Parker spiral model, and with a significant excursion into positive polarity bounded by PI5 and PI6. However, we note here that PI1, PI2, PI3 and PI4 are lacking in these predictions. This matter is discussed further in section \ref{timeseries_rss} where we recover these polarity inversions through tuning of the source surface height parameter.
	
	Examining the ensemble of models over different observations we see some scatter about the data, and in particular an unobserved ``bump'' in the field around 2018-11-4. This small disagreement with the data is mitigated by the time-integrated model which in all cases traces closer to the real data and ignores this unobserved bump. From this, we infer that the time evolution of the input magnetograms is important on the timescale over which the PSP observations took place but treating it as a time series of steady state models mitigates the lack of dynamics in PFSS modeling.
	
	Comparing the models with and without the $1/r^2$ scaling, we observe the negative interval around perihelion infers a near constant (flat) magnetic field at the source surface. This is consistent with the fact that this time interval corresponds to times when PSP was corotating or moving very slowly compared to the solar surface and so was likely crossing flux tubes very slowly and observing a very slow change in source region (see section \ref{fld_mappings}). 

	We also observe that using solar wind speed measurements to propagate field lines causes subtle improvement to model. For example, the timing of maximum field strength agrees better with data, the relative amplitude of the peak compared to the trough improves and the near perihelion field profile flattens in panel D vs B. The timing and steepness of the PI5 polarity inversion also improves.

	Meanwhile, PI6 demonstrates a limitation of PFSS: The prediction is for a smooth, protracted transition through Br = 0 whereas the data shows a sharp transition. MHD modeling \citep{Reville2017} has shown that further latitudinal evolution beyond the source surface sharpens the location of polarity inversions  by homogenizing the radial field in latitude, which better matches Ulysses observations \citep{Smith1995,Smith2011}. This could explain this discrepancy. 
	
	Additionally, even after time integration and correction for varying solar wind speed, the general negative field predicted prior to PI3 is approximately a factor of two weaker than measurements suggest at the source surface (Panel D). This indicates the same $\text{A}/r^2$ scaling (with a constant A) of the PFSS-derived field magnitude is not globally applicable to the whole time series, particularly as PSP gets closer to the polarity inversion line (PIL, see section \ref{timeseries_rss}) where PFSS predicts a drop off in field strength. The sharpening effect of non-radial expansion outside the source surface is a likely explanation for this discrepancy too as it predicts the field strength remains constant closer to the PIl than PFSS suggests.
	
	\subsection{Impact of variation of source surface height parameter}
	\label{SS}

	In this section, the impact on the results of varying the source surface height are investigated. While this is a numerical modeling parameter, it does have physical consequences on the predictions of the model affecting the total open flux, the apparent size of coronal holes and the complexity of the polarity inversion line which seeds the heliospheric current sheet in other models such as WSA \citep{Arge2003}.

	\subsubsection{Coronal Hole Distribution}

	\label{OPEN_FIELDLINES}

\begin{figure*}
		\centering
		\plotone{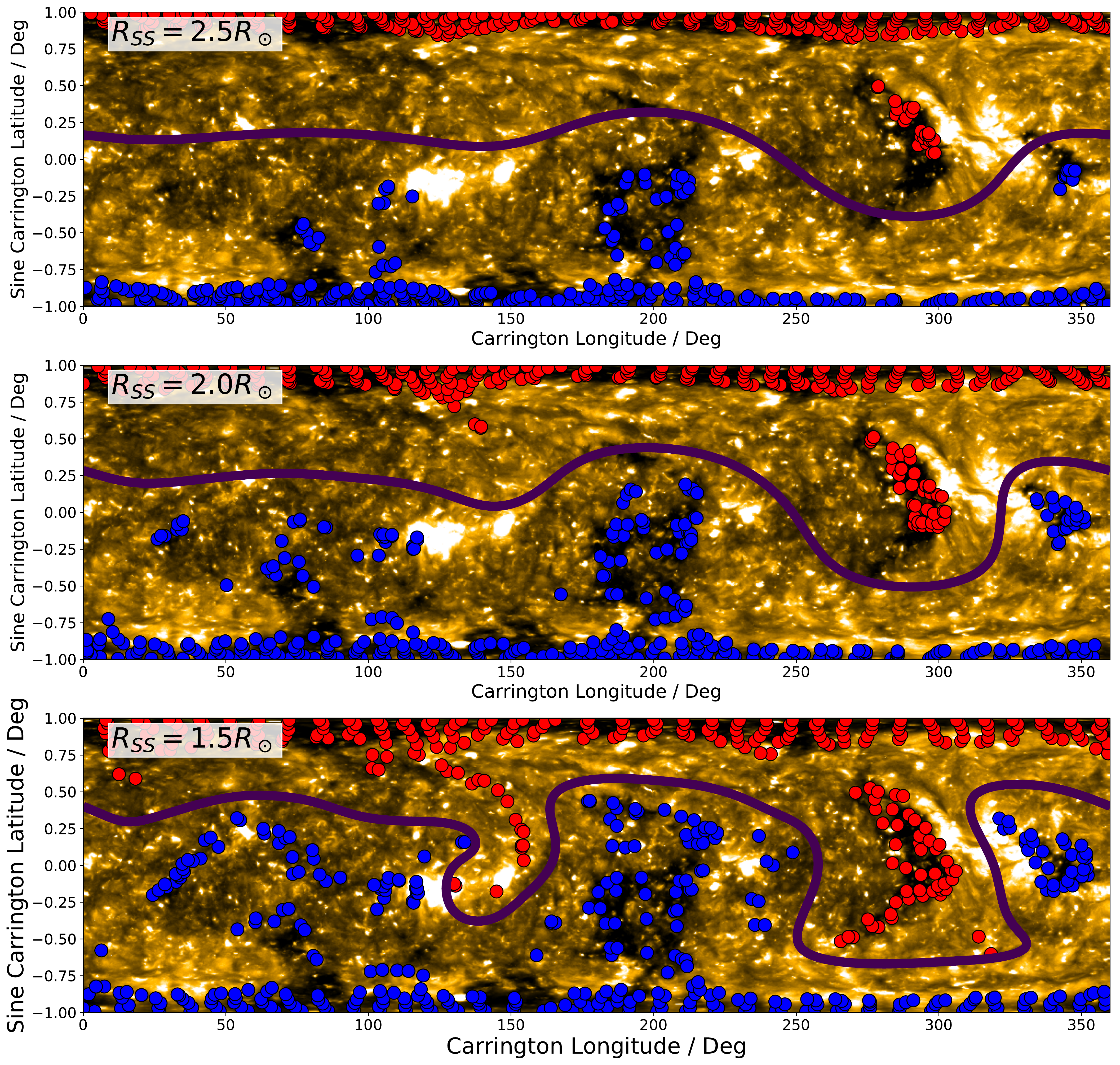}
		\caption{\textbf{Distribution of PFSS predicted open field line footpoints as a function of source surface height. } For the values of source surface height $2.5 R_\odot$, $2.0 R_\odot$,  $1.5 R_\odot$, field lines are initialized on a uniform grid at the source surface and mapped down to the photosphere. By definition, field lines initialized at the source surface are open and so this mapping shows the PFSS prediction of the source regions of open magnetic flux. For context, these mappings are overlayed on a synoptic map of the 193$\text{\AA}$  emission synthesized from STEREO/EUVI and SDO/AIA. At this wavelength, dark regions imply low density plasma regions in the lower solar corona, indicating the presence of coronal holes where open magnetic field lines allow plasma to evacuate into interplanetary space. The dark line overplotted is the PFSS neutral line at the source surface and can be seen to warp more at lower source surface heights. }
		\label{fig5}
	\end{figure*}

	In figure \ref{fig5} we begin our discussion of the source surface height parameter by comparing the footpoints of open field lines to synoptic maps of extreme ultraviolet (EUV) emission from the solar corona. This data is assembled from data observed during Carrington Rotation 2210 by STEREO/EUVI \citep{Wuelser2004} observing at 193$\text{\AA}$ and SDO/AIA \citep{Lemen2012} observing at 193$\text{\AA}$, these emissions are produced by Fe XII, a highly ionized state of Iron which is excited around 1,000,000 K. The brightness of these maps indicates the density of plasma at the  1,000,000 K isotherm which is approximately a surface of constant height in the lower corona. Dark regions identify coronal holes which are the probable locations of open field lines which allow plasma to escape outwards into the heliosphere resulting in an underdense region.
	
	By comparing the locations of footpoints of open field lines implied by the PFSS model, we can evaluate how accurately the model reproduces the observed coronal hole distribution. We use model results using the GONG 2018-11-6 map (centered on perihelion) and vary the source surface from 2.5 $R_\odot$ down to 1.5 $R_\odot$. The open footpoints are generated by initializing field lines from a uniform grid at the source surface where the field lines are all open by construction. We then use the \textit{pfsspy} field line tracing utility to propagate each line down into the model and find it's point of intersection with the lower boundary of the model.
	
	From comparing dark regions with the modeled footpoint locations we see a generally sensible solar minimum model result: the majority of open field lines map to the north and south polar coronal holes, with some equatorward extensions picked out along with individual isolated mid-latitude and equatorial coronal holes.
	
	However, we observe for the canonical $2.5 R_\odot$ that these isolated coronal holes are very underexpanded in the PFSS model, with particular emphasis on the small negative equatorial coronal hole around 340 degrees longitude and the more extended positive polarity one just below 300 degrees longitude. Lowering the source surface height to $2.0 R_\odot$ we find a more reasonable filling of these two features. Going further and examining a source surface height of $1.5 R_\odot$, we start to overexpand the  mid-latitude and equatorial coronal holes and produce some footpoints without obvious coronal hole correspondence.
	
    It should be noted a precise match between dark EUV images and the footpoints of open field lines is not expected since the EUV coronal hole boundary is somewhat wavelength dependent. Nevertheless, it appears 2.0 $R_\odot$ is a reasonable height for a globally consistent PFSS model with regard to observed coronal hole locations. This inference is consistent with the parameter chosen for the predictive modeling work for PSP E1 by \cite{Riley2019} to make PFSS results agree with MHD modeling as well as possible without overexpanding coronal holes. \citet{Lee2011} also found that the canonical 2.5 $R_\odot$ source surface height resulted in underexpanded coronal holes for a similarly quiet solar minimum carrington rotation. More recently, \citet{Nikolic2019} has made the same observation for PFSS extractions of GONG maps from 2006 to 2018. Further evidence for this lower source surface height is presented in appendix \ref{AppB} where we compute a cost function as a function of source surface height and show the ``optimum'' source surface height is significantly lower than 2.5$R_\odot$ for both GONG and ADAPT evaluations. We use these findings to justify our use of this parameter value in our comparison in figure \ref{fig4}, and to inform our further investigation of variation of this parameter.  

	\subsubsection{Impact on Timeseries Predictions}
	\label{timeseries_rss}

\begin{figure*}
		\centering
		\plotone{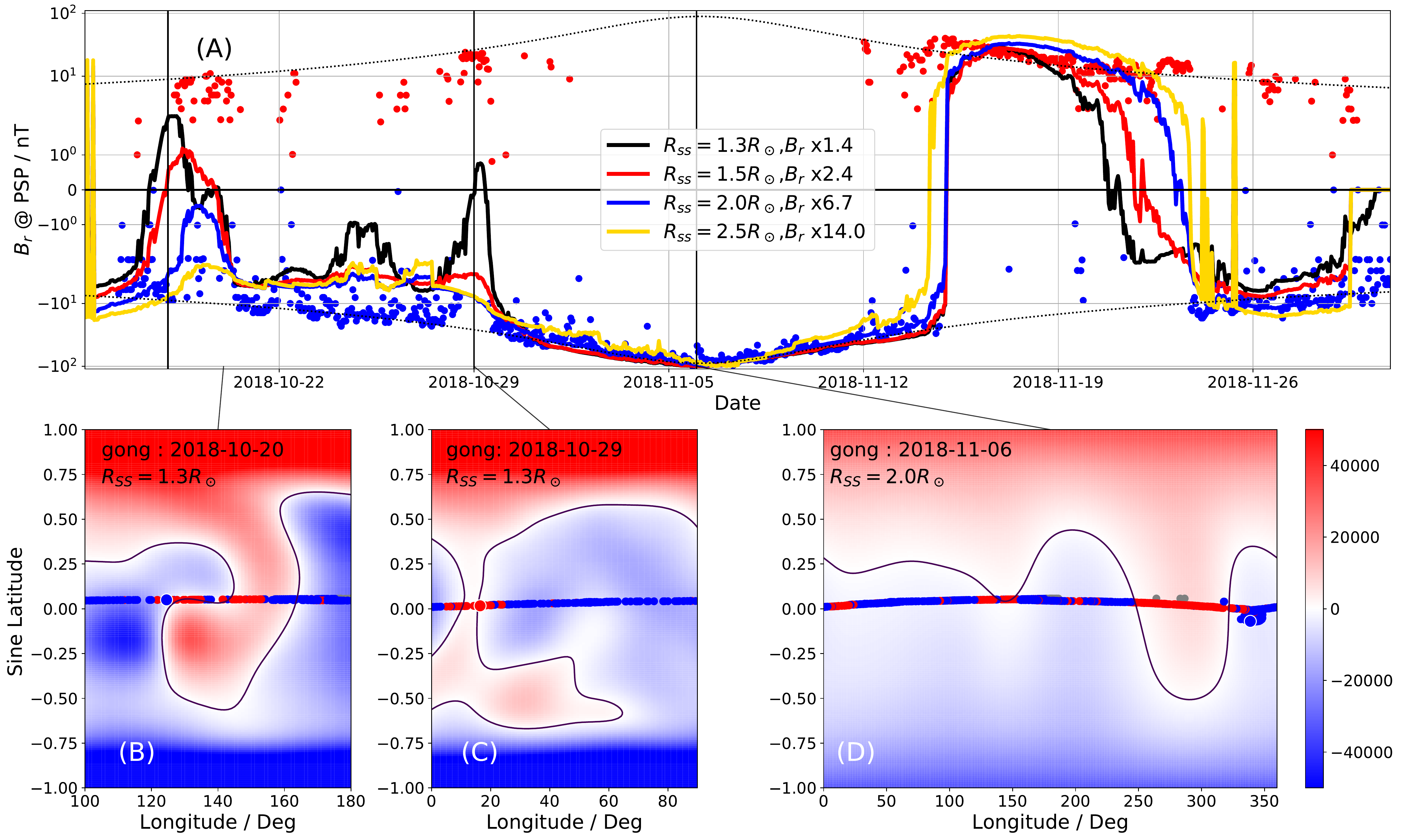}
		\caption{\textbf{PFSS Results as a function of Source Surface Height}. Panel (A) : comparisons of model predictions for different source surface heights. Predictions use time evolving magnetograms and SWEAP measurements. The scaling factors required to get the peak field strengths of the predictions and data to match for each source surface height are indicated in the legend. 1.3$R_\odot$ and 2.0$R_\odot$ models are as shown in \citet{Bale19} figure 1(b). Panels (B)-(D) show the magnetic field strength at the source surface predicted by PFSS projected onto longitude vs. sine(latitude). A black contour indicates the polarity inversion line. Overlayed on this are the source surface footpoints of PSP colored by the measured polarity. Panels (B) and (C) show zoom-ins on PIL structure consistent with small scale positive polarity inversions measured on 2018-10-18 and 2018-10-29 respectively for a source surface height of $1.3 R_\odot$. Panel (D) shows the map for a $R_{SS}=2.0 R_\odot$ model over the whole Sun. PSP starts at approximately 180 degrees longitude and tracks from right to left as time passes. The lines between panels indicate the time on the timeseries when the model shown was evaluated. }
		\label{fig6}
	\end{figure*}

	We next examine the results of varying the source surface height parameter on the time-integrated best fits from section \ref{RESULTS1}. The results are summarised in Figure \ref{fig6}. Panel (A) shows time series predictions for source surface heights ranging from $2.5 R_\odot$ down to $1.3 R_\odot$. The $B_r$ values are shown on a symmetric log scale to emphasize the polarity inversion features. The scaling factors applied to match the model peak field with the measured peak field are indicated in the legend. These range from 14.0 at the canonical $R_{SS}=2.5R_\odot$ down to order unity for $R_{SS}=1.3R_\odot$.   Panels (B-D) are shown to contextualize the time series. In each of these, a colormap of magnetic field strength at the source surface is plotted along with the polarity inversion line (contour of $B_r=0$) in black. On top of this, the PSP trajectory ballistically propagated down to the source surface is shown and colored by \textit{measured} polarity. Where the measured color matches the color of the source surface below it, the model and data are in agreement. Panels B and C show model results evaluated from GONG magnetograms on 2018-10-20 and 2018-10-29 respectively, both for the extremely low source surface height $1.3 R_\odot$ and zoom in to a specific part of the Sun to highlight specific PIL topology associated with the PI1-PI2 and PI3-PI4 intervals. Panel D meanwhile offers a more global view with the model at $2.0 R_\odot$ showing the entire encounter. PSP began the encounter at approximately 180 degrees longitude and tracks in the direction of decreasing longitude in time, approximately reaching the position it started at at the end of the time interval considered.

\begin{figure*}
		\centering
		\plotone{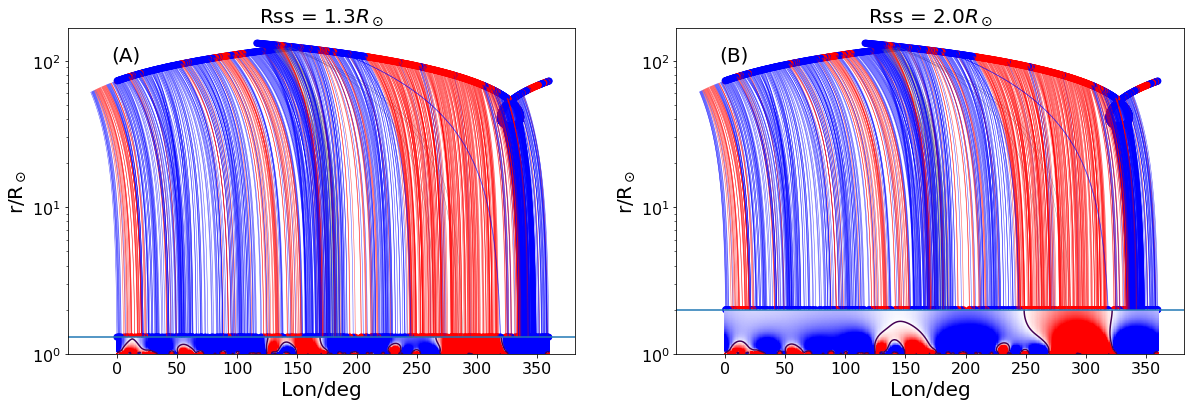}
		\caption{\textbf{Illustration of impact of Lowering SS Height on Model}. A comparison of the PFSS model and connection to PSP is shown in a plot of log(r) vs radius. Projecting everything into the ecliptic plane, we plot the PSP trajectory colored by measured polarity, then plot the parker spiral field lines down to the source surface, also colored by measurements. Below the source surface we plot a color map of an equatorial cut through the model colored by $B_r$, with black lines indicating contours of $B_r = 0 $. Panel (A) shows the results at $R_{SS} = 1.3 R_\odot$ while Panel (B) shows the results at $R_{SS} = 2.0 R_\odot$. Comparison between model and observations are made at the source surface.  }
		\label{fig7}
	\end{figure*}

	Examining each measured polarity inversion in turn we observe that the timing of PI5 is accurately predicted by all models regardless of the source surface height. PI6 is well predicted for $R_{SS} \ge 2.0 R_\odot$, but for lower $R_{SS}$ it is predicted to occur too early. This is discussed further with figure \ref{fig7}. 
	
	Conversely, PI1, PI2, PI3 and PI4 are entirely missed in predictions using $R_{SS} \ge 2.0 R_\odot$. Only at the very lowest source surface heights do these features convincingly appear in the prediction.
	
	Panel D shows the overall distibution of magnetic field at the source surface for $2.0 R_\odot$. An overall flat PIL generally skewed north of the equator explains why the near equatorial PSP mainly connected to negative polarity. A major southward warp of the PIL between 250 and 330 degrees longitude explains PSP's major excursion into positive polarity between PI5 and PI6 which is generally consistent with the shape of the HCS inferred by \citet{Szabo2020}. However the amplitude of the warps we infer and hence distance from the current sheet are larger than expected at the radius of PSP. This is expected since modeling beyond the source surface (either WSA's Schatten Current Sheet or flow dynamics in MHD) has the effect of flattening the PFSS-derived PIL as it evolves into the HCS (see fig 2. \citet{Szabo2020}). However, in spite of the discrepancy in distance to the current sheet, the timing of crossings and overall shape of the predicted magnetic timeseries is largely unaffected and remains consistent with other models.
	
	Panels B and C show that when the source surface is lowered to $1.3 R_\odot$ thin tenuous southward extrusions of the PIL develop at the correct longitudes at which PSP observed PI1-PI2 and PI3-PI4 intervals. However, to achieve this we find that the model must be generated using magnetograms from very close in time to when PSP was at that location. This suggests these features evolved quickly.
	
	In the case of the PI1-PI2 interval, and the PI2 transition in particular, the predictions at the source surface can be traced to a distinct photospheric feature: A simple dipolar active region lies directly below the longitude where PSP observed positive polarity. The implied connection is discussed further with figure \ref{fig8}(B).
	
	These model results and correspondence to data are compared in an alternative format in figure \ref{fig7}. Here, confining all data to the solar equatorial plane, we plot the PSP trajectory colored by measured polarity as log(radius) vs longitude. For each point along the trajectory we trace the Parker spiral used to connect it to the source surface and color this by data too. Below the source surface we plot a colormap of an equatorial cut through a PFSS model evaluated using GONG data from 2018-11-6 (perihelion). The comparison between model and observations is made at the source surface by comparing the model color just below and the observation color just above. By plotting log(radius) we are able to display both the interplanetary scale of the Parker spiral field lines and the coronal scale PFSS model. In panel (A) we plot the low source surface height model ($R_{SS}=1.3R_\odot$) , and Panel (B) shows the high source surface model ($R_{SS}=2.0R_\odot$). 
	
	Examining the evolution of the equatorial cut with radius, it can clearly be seen how higher order structures initialized at the photosphere smooth out and becomes simpler with increasing altitude. A general picture in the equatorial cut shown here is an overall negative polarity with small perturbations of positive polarity, most notably at 10, 55, 150 and 300 degrees longitude. Examining the high source surface case, panel (B), we see how only the 300 degrees longitude feature intersects the source surface and the others close over and revert back to negative. The 300 degrees longitude feature also exhibits expansion in it's longitudinal extent with altitude. On comparison to the observations, we notice this evolution is a vital element to explaining polarity inversion timings observed by PSP. 
	
	In comparison, panel (A) shows how the low source surface height is required for the ``low-lying'' strucures at 10 and 150 degrees longitude to be opened to interplanetary space. We also see how this source surface is likely the lowest possible since taking it further down would open up structures at 55 and 240 degrees longitude, which have no observational support. However, while the lower height model recovers these small structure better, it intersects the 300 degree feature before it has expanded sufficiently, and so produces a prediction of positive to negative polarity inversion much earlier than was observed (figure \ref{fig6}A). These competing model features make the limitations of a single height source surface apparent: with this constraint our ability to globally match all observed features is hindered.
	
	Figure \ref{fig7} is also useful in making a qualitative assessment of the uncertainty associated with ballistic propagation. According to \citet{Nolte1973}, this method has an estimated uncertainty in projected longitude of $\pm 10$ degrees.  Figure \ref{fig7} allows visualization of the extent to which a shift of 10 degrees could shift projected footpoints relative to the observed structure in the PFSS model. For example, the westward expansion of the 300 degrees feature from 1.3 to 2.0 $R_\odot$ is approximately 20 degrees and so a 10 degree shift of the observed transition from positive to negative polarity could not explain the discrepancy in the two source surface height models. The blue sector from 330-360 degrees longitude indicates the extent of the source surface region which maps down to the coronal hole we infer PSP was connected to at perihelion (section \ref{fld_mappings}). This is similarly larger in extent that the uncertainty and so we can state this connection is robust within uncertainty. With the smaller scale positive polarity features which are of order 10 degrees in extent a 10 degrees shift could just drive the footpoints of measured positive polarity out of consistency however it would not be sufficient to drive them to connect to any of the other positive polarity regions so the connectivity we infer remain the most likely sources of those measurements. Overall, we argue that 10 degrees can affect the details of the agreement of model and data but not our overall conclusions. In addition, with PSP orbiting much closer to the corona than the 1AU, this may result in a lower longitude error as the magnitude of the longitude correction is smaller. The excellent matching of observations and modeling to much less than 10 degrees may be evidence of this.
	
	\subsection{Implied Field Line Mappings}
	\label{fld_mappings}
	
	Having built confidence in the modeling approach by observing good time series predictions, as well as identifying it's response to variation in input parameters, we now explore the implications of the model results for the sources of the solar wind measured by PSP during it's first encounter. 
	
\begin{figure*}
		\centering
		\plotone{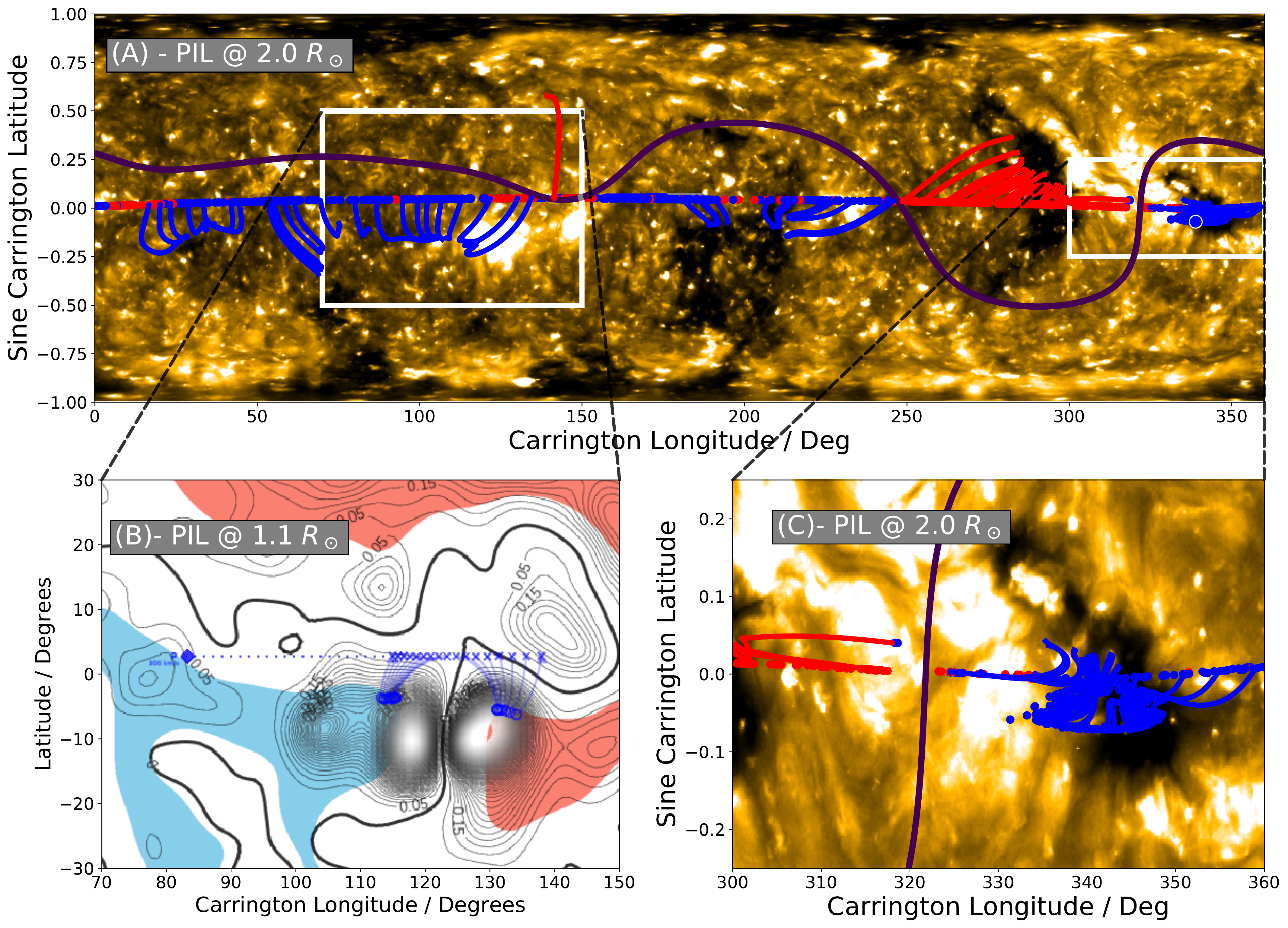}
		\caption{{PFSS Predictions of photospheric connectivity of PSP}. PFSS field line traces are initialized at the source surface footpoints of PSP and propagated down through the corona to the solar surface. These are contextualized with the same 193$\text{\AA}$ map from figure 5, and the polarity inversion line at the source surface from the model. Panel (A)  shows a synoptic view of the whole encounter with the model evaluated at $R_{SS}=2.0 R_\odot$. Panel (B) shows a zoom in to the 2018-10-20 polarity inversion with the model evaluated at $R_{SS}=1.4 R_\odot$ using the DeRosa/LMSAL surface flux transport model. Contours of $B^2$ are shown and the polarity inversion line evaluated at $1.1 R_\odot$ is plotted in bold. Coronal hole regions are shown as red and blue shading. The blue diamond at 85 degrees longitude indicates the carrington longitude of PSP, the crosses indicate the footpoints of PSP at the source surface. The circles indicate \textbf{magnetic} footpoints at \textbf{1.1 $R_\odot$} and the crosses and circles are connected by field line extrapolations. Panel (C) shows a zoom in of the connectivity around perihelion with $R_{SS}=2.0 R_\odot$. The field line mappings indicate connectivity to an equatorial negative-polarity coronal hole preceding a polarity inversion. Field lines shown in panels (A) and (C) are from same model evaluation depicted in \citet{Bale19} figure 1(c,d). }
		\label{fig8}
	\end{figure*}

	Similar to how open field footpoints were generated for figure \ref{fig5}, we use the PSP trajectory projected down to the source surface to initialize a series of fieldlines which we propagate down into the model corona using the \textit{pfsspy} field tracing routine. The results of this tracing with $R_{SS}=2.0R_\odot$ are summarised in figure \ref{fig8}A where we overplot the fieldlines colored by their polarity in the model on top of the same EUV data from figure \ref{fig5}. We also show the PSP trajectory projected on to the source surface and colored by \textit{measured} polarity as in figure \ref{fig6} (B,C,D). 
	
	In figure \ref{fig8}(B,C), we examine the field line traces of two subregions of the Sun of particular interest. Panel (B) shows results around the active region which appears associated with the PI1-PI2 interval. The modeling results shown here are using the DeRosa/LMSAL model evaluated on 2018-10-20 with a source surface height of $1.4R_\odot$, the highest value of $R_{SS}$ for which the model and measurements agree \citep[see ][]{Panasenco2020}. The faint contours of $|B|^2$ illustrate the topology of the active region, while the bold contour depicts the PIL at 1.1$R_\odot$, indicating the polarity structure at the base of the corona. Crosses mark the positions on the source surface PSP connects to and track from right to left in time. The circles indicate the field footpoints these map to. It can clearly be seen that in this time interval the footpoints jump from the positive to negative region of active region and that the PFSS-inferred neutral line is guided along the neutral line of this active region. It is difficult to argue PFSS accurately captures the magnetic topology around such non-potential structures, but it is compelling that the longitude of this region matches with the projected longitude of PSP at the same time it measured PI2 implying at the very least an association of the observed feature with the measurements. This fine tuning of the crossing timing and more modeling with the DeRosa/LMSAL model is expanded on in more detail in \citet{Panasenco2020} in which the authors explore optimizing source surface height for various polarity inversion case studies, suggest a resulting fitted non-spherical source surface height profile, and study the propagation and loss of Alfv\`enicity of slow wind streams en route to 1AU. 
	
	Panel (C) zooms in on the region of the Sun where PSP was located during the two periods of corotating either side of perihelion. The results here are shown for $R_{SS} = 2.0 R_\odot$ but the qualitative conclusions here are largely independent of this choice. The major implication here is that for the entire ``loop'' part of the trajectory PSP was connected very stably to the small equatorial coronal hole at around 340 degrees longitude. 	This suggests PSP was connected to a region of the Sun of less than 10 degrees angular extent for over two weeks, over which time it's radial distance varied significantly. Therefore data from this interval can be interpreted as a time series of evolving solar wind from a single source convolved with changing sampling radius. Panel (C) also shows striking agreement between the location of the PIL at approximately 320 degrees longitude and the point at which PSP measurements indicated a polarity inversion from negative to positive.
	
	Other general observations from figure \ref{fig8}A indicates that PSP connected to another (but this time positive) equatorial coronal shortly after perihelion, but with much more rapid footpoint motion tracking northwards over time. Comparison to SWEAP measurements suggests this configuration is responsible for the longest interval of fast wind observed in encounter 1 (figure \ref{fig1}B). After crossing back from positive to negative polarity on November 23, it continued to be connected to mid latitudes via a very large negative coronal hole.
	
	\section{Discussion}    
	\label{DISCUSSION}

	Overall, despite it's simplicity, our PFSS + ballistic propagation scheme produces compelling predictions of the $B_r$ timeseries PSP measured during it's first encounter. The large scale features were well reproduced including the flat source surface field convolved with $1/r^2$ variation, the predominantly negative polarity and the timing of significant excursions into positive polarity. The model results are likely distorted by unmodeled chromospheric currents at the interior to the model, as well as field re-organization beyond the source surface, however empirically it appears these distortions were not great enough to destroy this overall good agreement. The applicability of PFSS is likely enhanced by the extremely low activity on the Sun at the time of observations which would result in fewer strong current systems. The impact of latitudinal reorganization between the source surface and $10R_\odot$ \citep{Reville2017} is likely mitigated by the near equatorial sampling of the spacecraft. PSP's orbital inclination is $\sim 4$ degrees relative to the solar rotation axis which is even less than the Earth-ecliptic inclination of $\sim 7.25$ degrees. Even so, we do see PI6 is predicted to be a smoother transition than observations suggest and this sharpening can be explained by modeling beyond the source surface.
	
	We also found the time evolution of the input photospheric maps was important to take into account, resulting in more robust and accurate time series predictions compared to those generated with a single magnetogram. The slow angular velocity of PSP compared to the surface of the Sun around perihelion likely plays a role in this finding, meaning since PSP tracked very slowly around the Sun in the corotating frame, the photospheric magnetic field changed significantly in this time interval. Since PFSS is a time independent model it is not possible to include dynamics, however we have shown that this limitation can be mitigated by treating the model as a snapshot of assumed static coronal structure sampled by the spacecraft for a small timestep, and chaining together these snapshots to approximate dynamical evolution.
	
	Next, we discussed the impact of varying the source surface height parameter and found that on lowering it, new polarity inversion features emerged in the prediction which were consistent with observations. However, while this improved the predictions at some heliographic locations it worsened it at others, suggesting there is no clear ``best'' source surface height (see also Appendix \ref{AppB}). This highlights the limitation of having a spherically symmetric source surface boundary condition. This qualitative finding is unsurprising given the lack of apparent spherical symmetry in observations of the outer corona, and development of PFSS since it's inception have sought to generalize beyond a spherical boundary \citep[e.g.][]{Levine1982}. Comparisons to MHD modeling have suggested a boundary of near radial field which is far from spherical \citep[e.g.][]{Riley2006}. This latter result however generally showed the dominant perturbation to sphericity was a latitudinal effect at solar minimum with the source surface elongated at high latitudes into a prolate spheroid shape. Given PSP's close to equatorial orbit and limited connectivity to high latitudes, producing model results which agree with observations is unlikely to contain information about the solar magnetic field at high latitudes. Instead, our results here are suggestive of variation in the height with \textit{longitude} and perhaps with localized perturbations to the source surface below the canonical radius over specific magnetic structures as explored in \citet{Panasenco2020}. 
	
	The successful prediction of new observed features at low $R_{SS}$ implies small, short lived magnetically open structures which persist out to interplanetary space are normally smoothed out by PFSS modeling but can be recaptured by investigating lower source surface heights. Figure \ref{fig2} suggests such small scale features may even be measurable at 1AU. One of these features appeared associated with a dipolar active region with a pronounced neutral line. Active regions are typically highly non-potential and dynamic, requiring MHD or non-linear force free modeling approaches. \cite{Riley2006} found PFSS modeling (with the canonical $R_{SS}=2.5R_\odot$) lacking in modeling fields in the vicinity of such a feature. Nevertheless PSP observed a polarity inversion from positive to negative at exactly the time low source surface height PFSS modeling suggests connectivity would have switched from the positive to negative lobes of the active region (Figure \ref{fig8}B). This implies in spite of the non-potential details, PFSS can be useful in associating such structures with observations in the inner heliosphere. The possibility of open field lines connecting to active regions has been discussed before for example in \cite{Svestka1977}.
		
	A brief discussion of the scaling factor as a function of $R_{SS}$ shown in figure \ref{fig6} is also warranted. We note that for $R_{SS} \ge 2.0$ a $1/r^2$ scaling from the source surface out into the heliosphere produces under-predicts the in situ field strength by an order of magnitude. This was first observed early in the history of PFSS by \cite{Levine1977}. A simple interpretation is that non-radial evolution beyond the source surface is important in connecting the source surface far out to the heliosphere. At the source surface, PFSS predicts stronger field at the poles than the equator and it is known from Ulysses observations \citep{Smith1995,Smith2011} that this relaxes to a latitudinally independent state further out. The simplest configuration for this to happen is for the polar field to decrease faster than $1/r^2$ and for the equatorial field to drop more slowly than $1/r^2$ which is consistent with an underprediction of the field at low solar latitudes. It is possible that this underprediction could be mitigated with a Schatten Current Sheet model \citep{Schatten1972} as used in the WSA model, and this will be the subject of future work. However, from the results plotted in figure \ref{fig4}, the overall shape of the $B_r$ time series is consistent to within this scaling factor from which we conclude this ``extra'' field strength is only weakly dependent on solar longitude, with some worsening agreement at higher PSP radii (early and late in the encounter). That the scaling factor drops to order unity with decreasing source surface height demonstrates how the radial field inside the PFSS model domain drops faster than $1/r^2$ since the dominant component is a dipolar $1/r^3$ field. Flux which opens up to the solar wind very low down in the corona likely escapes purely radially, which may explain why $1/r^2$ predicts the correct magnitude at PSP for these cases.
	
	Lastly, in addition to the active region connectivity discussed above, we presented a synoptic view of the connectivity implied by PFSS during the whole encounter. The relatively low $2.0 R_\odot$ source surface height suggests predominantly equatorial and mid-latitude connectivity as opposed to deep within polar coronal holes. This is consistent with PSP's predominantly slow wind observations (Figure \ref{fig1}B). However we note here that raising the source surface height to it's canonical value can change the connectivity very early in the encounter to the southern (negative) polar coronal hole. Given the time interval examined is dominated by the perihelion loop, our inference of a ``best'' source surface height is likely skewed towards the corotation interval and so statements of connectivity earlier or later in the encounter at $2.0R_{\odot}$ may be weakened. For example, \citet{Reville2020} see polar connectivity at intervals prior to October 29 with MHD modeling results, while \citet{Szabo2020} examine an ensemble of different model and establish PSP was very close to the HCS at this time and thus may be observing streamer belt plasma.
	
	Of particular interest is the connectivity in the 2 week interval surround perihelion (figure \ref{fig8}C) which shows throughout the corotational period PSP was magnetically connected to an isolated, negative polarity equatorial coronal hole. This was also predicted prior to the encounter by \citet{Riley2019} and now appears well corroborated by \citet{Kim2020,Reville2020,Szabo2020}. The coronal hole is approximately 10 degrees in longitudinal extent at the 193$\text{\AA}$ isosurface which implies a linear distance scale of order 100 Mm. Therefore without assuming detailed knowledge of the PFSS-derived footpoints we can state the source region for the sampled solar wind from this 2 week period (2018-10-30 to 2018-11-14) was confined to this 100 Mm size region, and thus variation in in situ data from this time can be interpreted as measurements of dynamics of the solar wind emitted by an approximately fixed source on the Sun \citep[see e.g.][]{Bale19}. This connectivity will also lead to interesting discussion on sources of the slow solar wind which is very much an active area of research \citep[e.g.][]{Wang2019}. 
	
	In summary, we have presented a first attempt at global coronal and inner heliospheric modeling to contextualize observations made by PSP in it's historic first solar encounter. Our potential field based modeling scheme is extremely simplistic and it will be vital to make detailed further comparison with concurrent or future global modeling work e.g. MHD \citep{Kim2020 Reville2020} or other PFSS models \citep{Panasenco2020,Szabo2020}. Nevertheless we report various pieces of evidence that suggest the limitations of our modeling have been mitigated sufficiently to claim real and useful contextual information of PSP's magnetic connectivity. In particular, we find PSP spent the 14 days surrounding perihelion connected to a small negative equatorial coronal hole and may have also sampled open flux tubes associated with an active region around 2018-10-20 prior to perihelion.
	
	The data discussed in this work comes from the first of many planned encounters for PSP. At the time of writing, encounters 2 and 3, which both sample down to the same perihelion distance of 35.7 $R_\odot$, have taken place. Following encounter 3, PSP will perform a gravity assist with Venus and begin to probe deeper into the solar atmosphere eventually reaching below $10 R_\odot$. Repeating the analysis of this work on these future encounters will be interesting and will, for example, allow us to control for the impacts of solar activity on modeling results.

	\section*{Acknowledgments}
	
	{Parker Solar Probe was designed, built, and is now operated by the Johns Hopkins Applied Physics Laboratory as part of NASA’s Living with a Star (LWS) program (contract NNN06AA01C). Support from the LWS management and technical team has played a critical role in the success of the Parker Solar Probe mission. The FIELDS and SWEAP experiments on the Parker Solar Probe spacecraft was designed and developed under NASA contract NNN06AA01C. The authors acknowledge the extraordinary contributions of the Parker Solar Probe mission operations and spacecraft engineering teams at the Johns Hopkins University Applied Physics Laboratory. S.D.B. acknowledges the support of the Leverhulme Trust Visiting Professorship program. S.T.B. was supported by NASA Headquarters under the NASA Earth and Space Science Fellowship Program Grant 80NSSC18K1201. This work utilizes data obtained by the Global Oscillation Network Group (GONG) Program, managed by the National Solar Observatory, which is operated by AURA, Inc. under a cooperative agreement with the National Science Foundation. The data were acquired by instruments operated by the Big Bear Solar Observatory, High Altitude Observatory, Learmonth Solar Observatory, Udaipur Solar Observatory, Instituto de Astrofísica de Canarias, and Cerro Tololo Interamerican Observatory. This work utilizes data produced collaboratively between Air Force Research Laboratory (AFRL) \& the National Solar Observatory (NSO). The ADAPT model development is supported by AFRL. The input data utilized by ADAPT is obtained by NSO/NISP (NSO Integrated Synoptic Program). This research made use of HelioPy, a community-developed Python package for space physics \citep{Stansby2019c}. This research has made use of SunPy v0.9.6 (\cite{Mumford2019}), an open-source and free community-developed solar data analysis Python package \cite{Sunpy2015}.}

	\appendix
	\section{Choice of Magnetogram Source}
	\label{AppA}
	
	As mentioned in section \ref{MODELING_METHOD}, although we show throughout this paper results using GONG zero point corrected magnetograms we did consider a range of other possible sources. Here we show that our results are largely independent of the choice but that GONG produces the most compelling predictions.

	\begin{figure*}
		\centering
		\plotone{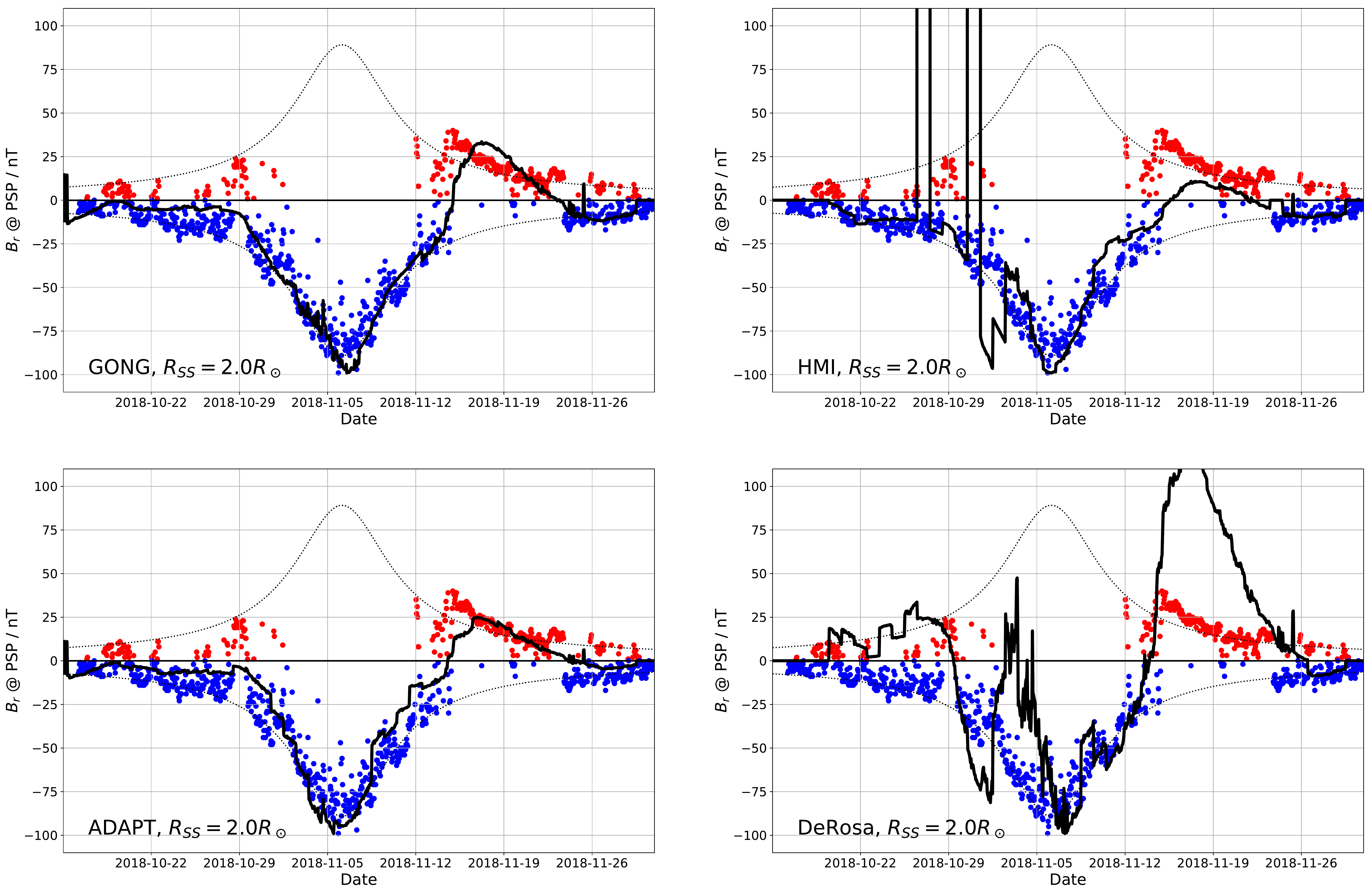}
		\caption{\textbf{Comparison of timeseries predictions using different magnetograms}. The ``best'' GONG timeseries prediction from figure \ref{fig4} (main text) is shown here in comparison to the same procedure applied to magnetograms from HMI, ADAPT and the DeRosa/LMSAL model. GONG produces the smoothest prediction on time integration but the general picture of negative polarity, $1/r^2$ variation and the times of polarity inversions are not strongly perturbed by choice of magnetogram.}
		\label{appfig}
	\end{figure*}	

	\begin{figure*}
		\centering
		\plotone{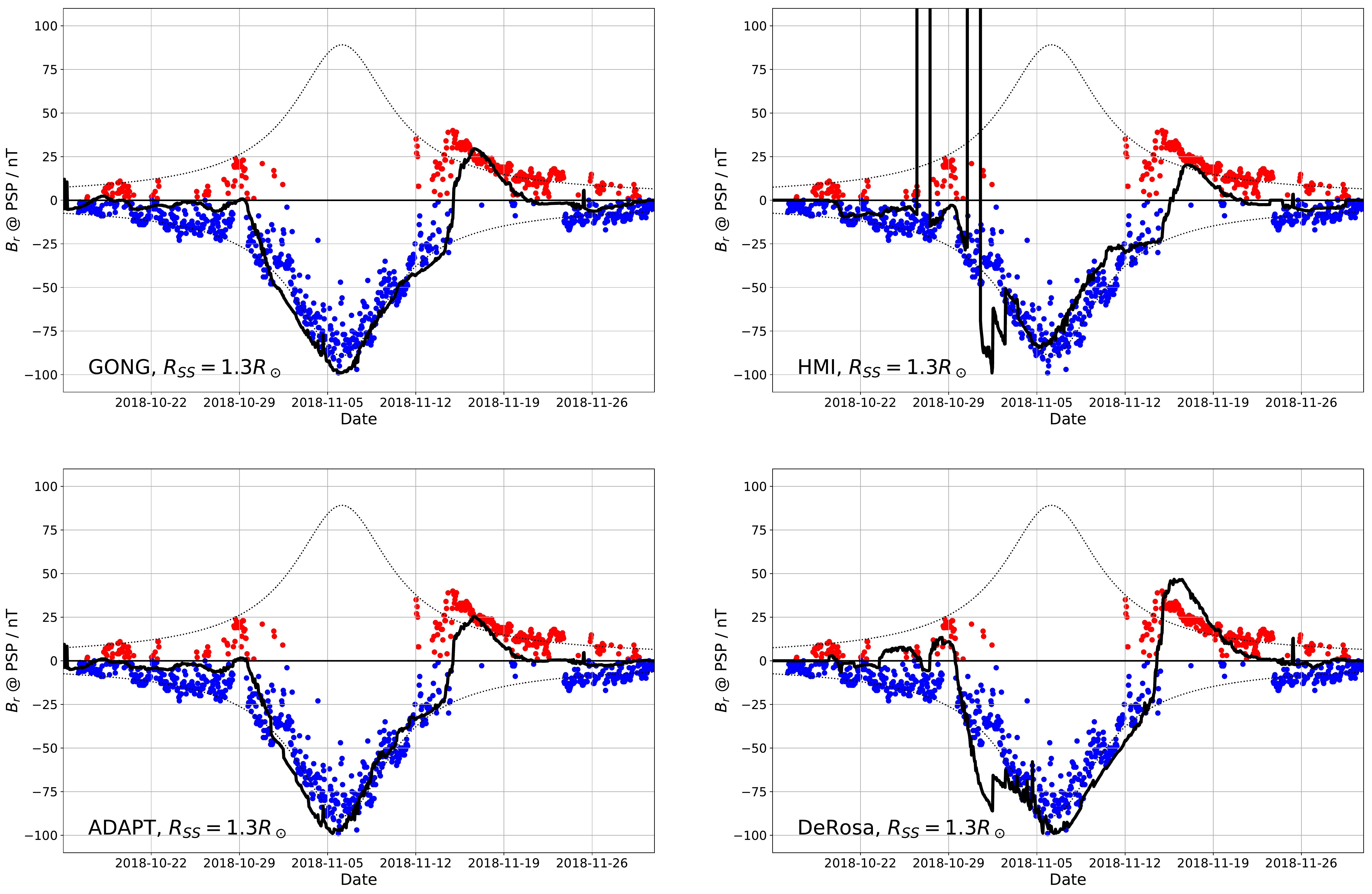}
		\caption{\textbf{Comparison of timeseries predictions using different magnetograms at low source surface height}. Compared to $2.0 R_\odot$, here all the models are very consistent with each other including LMSAL/DeRosa}
		\label{appfig2}
	\end{figure*}
	
	In figure \ref{appfig} we compare the GONG results from the main text figure \ref{fig4} with those obtained by an identical procedure on different magnetogram outputs. On the top row, are results from GONG and HMI which are both purely observational data products. Below are results from ADAPT and the DeRosa/LMSAL models which are surface flux transport models which assimilate the GONG and HMI data respectively. On visual comparison, we see GONG and ADAPT results are very similar as expected. HMI is largely consistent but with a too small a peak amplitude predicted, and has some spurious data from some of the magnetograms considered. The DeRosa/LMSAL-based result shows significant departure from the data: The period before 2018-10-29 is predicted to connect to the positive side of the heliospheric current sheet (HCS), the 2018-11-4 bump which is smoothed out by time integration (section \ref{RESULTS1}; main text) in the other models is still prevalent and the peak positive amplitude is significantly overestimated. The variation between these models demonstrates that because PSP traversed generally very close to the HCS during Encounter 1, the predictions are quite sensitive and a small change in the modeled PIL can produce a sudden reversal in polarity measured at PSP.  We err towards the GONG based data due to this empirical observation.

	In terms of the difference between GONG and ADAPT, the main noticeable change is that on time integration, the ADAPT prediction becomes ``choppier'', subsequent 3 day intervals don't smoothly meet each other. This is likely due to the extra physical modeling in ADAPT meaning flux variations are captured at higher time resolution than with pure GONG data. Nevertheless the major conclusions from section \ref{RESULTS1} are unchanged from use of either of these magnetograms. Since these fluctuations don't immediately appear to correspond to data we infer although they may be physical, they are likely smoothed out exterior to the source surface via processes not considered by PFSS. Beyond this, the choice of GONG vs. ADAPT does not affect the outcome of this paper and hence we make the choice of the smoother predictions and proceed with our analysis using this.
	
	For further comparison and to offer some insight into possible sources of discrepancy above, in figure \ref{appfig2} we compare the same choices of magnetograms with a source surface height of $1.3R_\odot$ which, as discussed in section \ref{timeseries_rss}, produces predictions of new small scale polarity inversions prior to perihelion.
	
	In this case we observe excellent consistency between all models including the DeRosa/LMSAL model. Since at this source surface height the polarity and field strength at the source surface is much more related to the field strength radially below, this is suggestive that high latitude field is the dominant cause of disparity in figure \ref{appfig} above, for example differences in how the unobserved polar regions are modeled. Nevertheless, we again note conclusions based on the GONG prediction are unchanged with different magnetogram sources and therefore work with this data in the main text.
	
	\section{Cost Function for Comparing Source Surface Heights}
	\label{AppB}
\begin{figure*}
	\centering
	\plotone{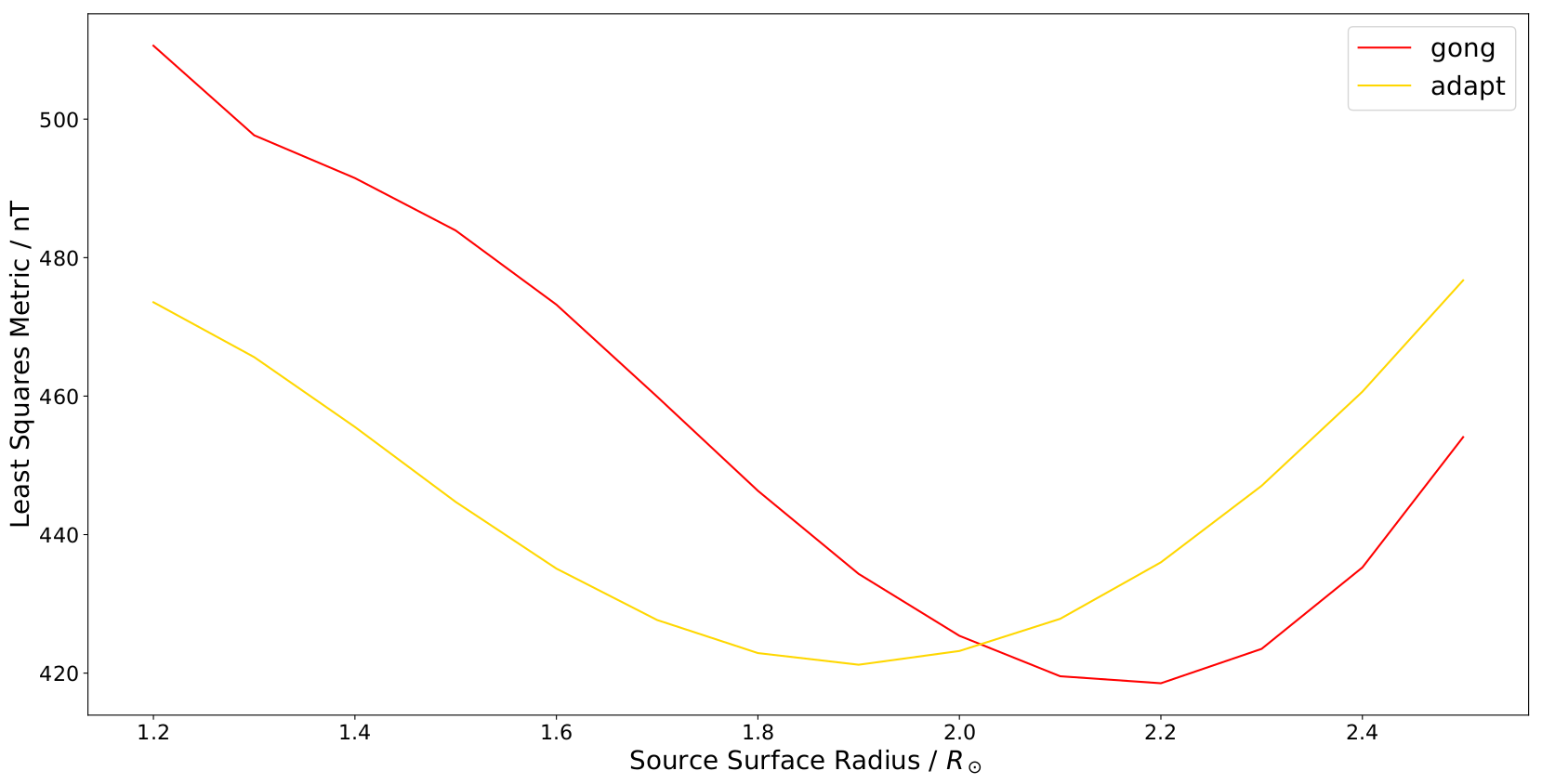}
	\caption{\textbf{Least squares metric computed between time integrated model results and PSP data as a function of source surface height}. Results for GONG and ADAPT are shown and exhibit a distinct minimum below the canonical 2.5$R_\odot$ for both cases. }
	\label{appfig3}
	\end{figure*}

	Further supporting evidence for our general use source surface heights below the canonical value of $2.5 R_\odot$ is shown in figure \ref{appfig3}. Here, we compute a least squares cost function evaluating the relative similarity between time integrated models (see section \ref{RESULTS1}).

	This cost function is given by :
	
	\begin{equation}
	L(M,O) = \sqrt{\sum_{i=0}^N (M_i - O_i)^2}
	\end{equation} 
	
	where the model, M and observations, O are N dimensional vectors. Since both the model and observations are expressed in nT this is also the unit of the cost function. In figure \ref{appfig3} we show the least squares result as a function of source surface height using the GONG (red) and ADAPT (gold) input magnetograms.
	
	Both models show a distinct minimum (best fit) at a significantly lower source surface height that $2.5 R_\odot$. The ADAPT ``best'' height is approximately 1.9$R_\odot$ while for GONG it is approximately 2.2$R_\odot$. However, these minima are both very broad and have overlapping full width half maxima. In addition, as discussed in the main text the concept of a single source surface height to fit all longitudes and for a 6 week long interval is likely not a good approximation. We settle on a value of $2.0 R_\odot$ to discuss a global picture in the main text, but note here a range of $\pm0.2 R_\odot$ will have very little affect on the overall goodness of fit. 
	

\end{document}